\documentclass{article}

\usepackage{iclr2027_conference,times}
\iclrfinalcopy

\usepackage[utf8]{inputenc} %
\usepackage[T1]{fontenc}    %
\usepackage{hyperref}       %
\usepackage{url}            %
\usepackage{booktabs}       %
\usepackage{amsfonts}       %
\usepackage{nicefrac}       %
\usepackage{microtype}      %
\usepackage{xcolor}         %
\usepackage{graphicx}
\usepackage{amsmath}
\usepackage{siunitx}
\usepackage{fvextra}
\usepackage{wrapfig}
\usepackage{float}
\usepackage{tikz}
\usetikzlibrary{arrows.meta,positioning}
\usepackage{cleveref}

\title{SMEBench: Benchmarking Neural Music\\Editing with Synthesized MIDI Edits}

\author{%
\parbox{\dimexpr\textwidth-2\tabcolsep\relax}{\centering
Matthieu Cervera$^{1,4}$, Francesco Paissan$^{1,2}$\\[0.3em]
Micael Antunes$^{3}$, Mirco Ravanelli$^{1,4}$, Ricard Marxer$^{1,5,6}$, Cem Subakan$^{1,2}$\\[0.6em]
\normalfont\small
$^{1}$Mila -- Quebec Artificial Intelligence Institute\\
$^{2}$Universit\'e Laval\\
$^{3}$National University of Ireland Maynooth\\
$^{4}$Concordia University\\
$^{5}$Univ Toulon, Aix Marseille Univ, CNRS, LIS\\
$^{6}$CNRS, ILLS
}%
}

\begin{document}

\maketitle

\begin{abstract}
Neural music editing requires a system to perform a requested change while
preserving the musical content that should remain unchanged. Existing
evaluations often lack a paired target edit, making these requirements difficult
to measure separately. We introduce SMEBench, a large-scale benchmark of
controlled music edits constructed by modifying symbolic MIDI performances and
rendering matched source and target audio. Each example contains a source, an
instruction, a paired target, edit metadata, and complementary descriptions.
The benchmark covers twelve instrumentation, articulation, dynamics, and pitch
operations and includes a balanced 3,600-example test set together with a
smaller real-audio complement derived from multitrack recordings. The paired
targets allow source preservation, target agreement, and textual compatibility
to be evaluated independently. We use these references to diagnose common
audio-editing metrics and benchmark six representative editing systems,
showing that the metric used to rank a system can materially change the
conclusion. The dataset is available at our repository\footnote{\url{https://huggingface.co/datasets/anonymous-account-submission/SME-Bench}}.
\end{abstract}

\section{Introduction}

Modern generative audio models learn powerful priors over musical structure
and acoustics
\citep{copet2024simplecontrollablemusicgeneration,
agostinelli2023musiclmgeneratingmusictext,
liu2024audioldm2learningholistic}. Recent work has begun to repurpose these
models for editing existing audio, where generation is constrained by both a
source recording and an instruction, usually provided in textual form. This creates a distinct control problem where the system must alter the original audio as requested without disturbing everything that should remain unchanged. A successful edit does more than produce plausible audio or closely resemble its source: it carries out the intended change, preserves the unaffected musical content, and introduces no unintended modifications.
\begin{figure}[t]
    \centering
    \resizebox{0.95\textwidth}{!}{
        \includegraphics[width=\textwidth, trim={1cm 1cm 3.1cm 5.1cm}, clip]{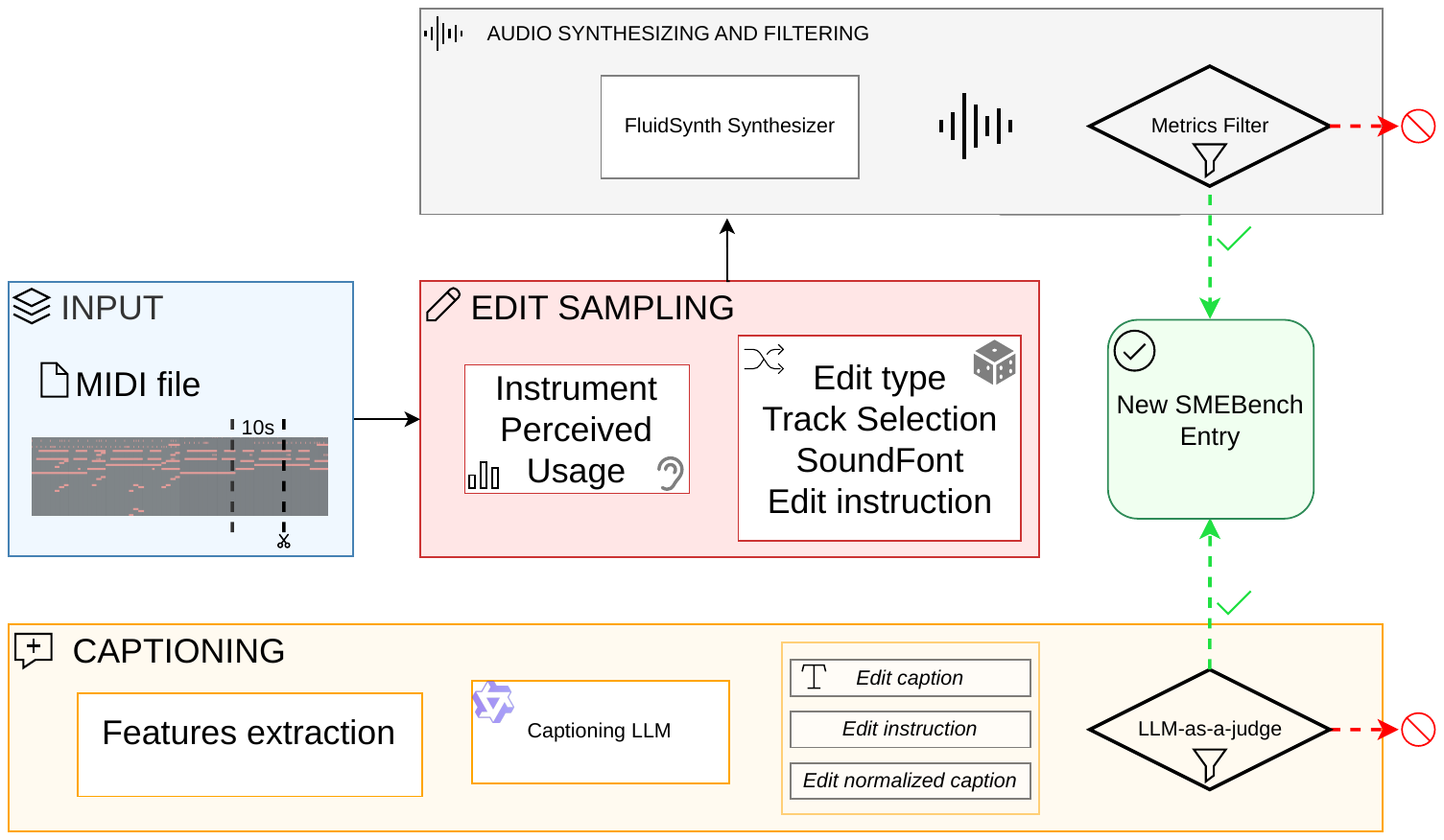}
    }
    \caption{\textbf{Overview of the SMEBench construction recipe.} A 10-second MIDI excerpt is analyzed to determine which edits and targets are applicable. The selected intervention is applied to the symbolic performance, after which the source and target are rendered with the same SoundFont and filtered for audibility and quality. Structured facts extracted from the MIDI, audio, and edit metadata ground four complementary textual descriptions. Examples that pass the audio and caption checks become benchmark entries containing a known intervention and a paired reference target.}
    \label{fig:smebench_pipeline}
    \vspace{-0.6cm}
\end{figure}
Evaluating these requirements is difficult when the intended edited audio is
unknown. Recent music-editing evaluations commonly rely on source audio and a
text instruction or target description without a paired reference edit
\citep{manor2024zeroshotunsupervisedtextbasedaudio,
cervera2025virtualconsistencyaudioediting}. In this
setting, source-audio similarity can measure preservation but may reward a
system that makes no edit at all. Likewise, text--audio alignment can measure
broad semantic compatibility while remaining insensitive to a local change in
instrumentation, articulation, or dynamics. Audio quality alone cannot
determine whether the correct transformation occurred. Consequently, the
metrics commonly adapted from audio generation conflate three different
properties: source preservation, text compatibility, and edit
correctness. Relying on these might rank systems without revealing what they successfully
changed.

This ambiguity has motivated edit-specific datasets and evaluation protocols
in computer vision \citep{zhao2024ultraeditinstructionbasedfinegrainedimage,
basu2023editvalbenchmarkingdiffusionbased}, but remains largely unresolved for
audio, specifically music. The central missing object is a paired target. That is, we would like to have access to an audio example that represents the requested transformation while holding the underlying
musical material fixed. Such targets serve as controlled references, allowing
preservation of the original material and agreement with the intended edit to
be measured separately. They also enable a direct test of whether an automatic
metric favors the intended target over an unchanged source, rather than merely
rewarding preservation or broad textual compatibility.

We introduce \textbf{SMEBench}, a benchmark for controlled music editing with
paired source and target audio. SMEBench uses MIDI as an intervention layer:
we apply known edits to symbolic performances and synthesize the source and
edited versions under matched conditions. Each example includes the source
audio, an edit instruction, the corresponding reference-edited audio, explicit
edit metadata, and complementary textual descriptions. This construction
provides controlled counterfactual pairs that differ by a known musical
intervention, allowing generated outputs to be compared both with the material
they should preserve and with the target they should approach. The current
release focuses on 10-second excerpts spanning a defined set of
instrumentation, articulation, dynamics, and pitch edits. Although the
controlled soundfont rendering enables precise, scalable interventions, it
also introduces a synthetic-to-real gap. To assess the extent to which our
conclusions persist under more realistic acoustic conditions, we additionally
construct a smaller paired subset from real multitrack recordings and report
results on it separately.

Using these paired references, we study not only how existing editing systems
perform, but also what their evaluation metrics measure. Identity and ground-truth anchors expose the different endpoints of the task. Returning the source gives perfect preservation without executing the edit, whereas the reference target represents the known intended transformation. Across common
audio--audio and text--audio metrics, our analysis shows that source
preservation, target agreement, and textual compatibility yield different and
sometimes conflicting assessments of the same output. These results motivate
paired, reference-based evaluation as a diagnostic complement to realistic
but unpaired music-editing benchmarks.

Our contributions are as follows:
\begin{itemize}

    \item We introduce \textbf{SMEBench}, a controlled benchmark containing 11,160 paired
    source and reference-edited audio, edit instructions, edit metadata, and
    complementary textual descriptions. Its balanced test set contains 3,600
    examples across twelve edit types, together with a real-audio complement
    composed of 416 pairs. We release the construction code as an extensible
    recipe for generating controlled, paired music edits.

    \item We use paired targets and interpretable reference anchors to
    characterize the behavior of commonly used audio--audio and text--audio
    metrics, showing that a system can score well for preserving the source or
    matching the text without being the closest system to the intended target.

    \item We benchmark SDEdit~\citep{meng2022sdeditguidedimagesynthesis}, DDIM
    inversion~\citep{song2022denoisingdiffusionimplicitmodels},
    ZETA~\citep{manor2024zeroshotunsupervisedtextbasedaudio}, Virtual
    Consistency Inversion~\citep{cervera2025virtualconsistencyaudioediting},
    SteerMusic~\citep{niu2025steermusicenhancedmusicalconsistency}, and
    MusicGen~\citep{copet2024simplecontrollablemusicgeneration}, and evaluate
    their outputs separately against the source and the intended target.
\end{itemize}

\section{Related Work}

Few datasets and evaluation sets target audio editing specifically. Some pair inputs and outputs, but derive targets through signal-level operations such as mixing, separation, masking, inpainting, or reconstruction; others provide prompts and annotations for evaluating music editors without target audio. This leaves a need for a common benchmark with paired targets produced through explicit interventions on musical content.

\paragraph{Paired audio-editing data.}
AUDIT~\citep{wang2023auditaudioeditingfollowing} constructs instruction-based
input--output triplets for general audio transformations, including addition,
removal, replacement, inpainting, and super-resolution. These triplets provide
paired supervision, but their targets are largely obtained through operations
such as mixing, separation, masking, and reconstruction. They do not intervene
on the underlying musical performance or expose a structured set of musical
changes. InstructME
\citep{han2023instructmeinstructionguidedmusic} similarly introduces an
instruction-guided music editing and remixing framework trained on a collected
music dataset. It considers operations such as adding, removing, extracting,
replacing, and remixing musical content, with targets constructed primarily
through stem-level manipulation. These datasets establish the value of paired
supervision, but cover a different setting from controlled interventions on
instrumentation and performance attributes in the underlying musical
representation. SMEBench instead modifies the symbolic performance before
rendering, producing paired targets with known changes to instrumentation,
articulation, dynamics, and pitch while holding the remaining musical content
fixed.

\paragraph{Evaluation sets without paired target audio.}
MedleyMDPrompts \citep{manor2024zeroshotunsupervisedtextbasedaudio} is a prompt-based
evaluation set constructed from the MusicDelta subset of
MedleyDB~\citep{bittner2016medleydb}. It contains real music recordings with
manually written source and target prompts, but does not provide the
corresponding reference-edited audio. ZoME-Bench
\citep{liu2025mediczeroshotmusicediting} selects music-editing examples from
MusicCaps~\citep{agostinelli2023musiclmgeneratingmusictext} and provides source
prompts, target prompts, and annotations across multiple edit categories, but
likewise lacks explicit target audio references. These resources offer greater
acoustic realism than synthesized MIDI and support evaluation on naturally
recorded music. Their unpaired design, however, means that edit execution must
be inferred from text alignment or changes relative to the source rather than
measured against a known edited outcome. SMEBench complements these realistic
evaluation sets with controlled paired references, making it possible to
compare an output separately with the source material it should preserve and
the intended target it should approach.

We review the main families of neural audio-editing methods and commonly used
evaluation metrics in Appendix~\ref{app:editing_pipelines_rw}.

\section{SMEBench: A Recipe for Controlled Paired Music Edits}
\label{sec:smebench}

SMEBench is a dataset and benchmark of source and target audio related by
known musical interventions. We construct it using the pipeline summarized in
Figure~\ref{fig:smebench_pipeline}, which proceeds from edit sampling to audio
rendering and textual grounding. We describe edit construction in
Section~\ref{sec:edit_construction}, caption generation in
Section~\ref{sec:captions}, and the benchmark composition, including a complementary evaluation set
constructed from real multitrack recordings, in Section~\ref{sec:benchmark_composition}.

\subsection{Edit Construction}
\label{sec:edit_construction}

We build SMEBench from MIDICaps~\citep{Melechovsky2024}, which contains
168,385 MIDI files with descriptive captions and extracted musical features.
MIDI provides access to the notes, timing, velocity, and instrument assignment
of a performance, allowing us to modify a selected musical attribute while
retaining the remaining symbolic content. For each example, we crop a random
10-second excerpt and inspect its active instruments and drums. This analysis
determines which edits are applicable. For instance, an instrument can only be
removed if it is present, and an extraction is discarded when it would leave
an empty or nearly empty target. Restricting the candidate operations in this
way reduces ghost edits, in which the requested change affects absent or barely
audible content, and degenerate edits that eliminate most of the excerpt.

\begin{figure}[t]
    \centering
    \resizebox{\textwidth}{!}{%
    \begin{tikzpicture}[
        >=Latex,
        line width=0.7pt,
        every node/.style={font=\small, align=center},
        root/.style={
            draw=black!65,
            fill=black!6,
            rounded corners=2pt,
            minimum width=2.4cm,
            minimum height=1.0cm,
            font=\small\bfseries
        },
        family/.style={
            draw=orange!70!black,
            fill=orange!10,
            rounded corners=2pt,
            minimum width=2.7cm,
            minimum height=0.8cm,
            font=\small\bfseries
        },
        subgroup/.style={
            draw=blue!60!black,
            fill=blue!6,
            rounded corners=2pt,
            minimum width=2.2cm,
            minimum height=0.72cm,
            font=\small\bfseries
        },
        operations/.style={
            draw=black!45,
            fill=white,
            rounded corners=2pt,
            minimum height=0.72cm,
            inner xsep=7pt
        },
        edge/.style={draw=black!55, -{Latex[length=2.2mm]}}
    ]
        \node[root] (root) {SMEBench edits};

        \node[family, right=1.4cm of root, yshift=1.95cm]
            (instrumentation) {Instrumentation};
        \node[family, right=1.4cm of root, yshift=0.65cm]
            (articulation) {Articulation};
        \node[family, right=1.4cm of root, yshift=-0.65cm]
            (dynamics) {Dynamics};
        \node[family, right=1.4cm of root, yshift=-1.95cm]
            (pitch) {Pitch};

        \node[subgroup, right=1.25cm of instrumentation, yshift=0.48cm]
            (instrument) {Instrument};
        \node[subgroup, right=1.25cm of instrumentation, yshift=-0.48cm]
            (drums) {Drums};

        \node[operations, right=1.15cm of instrument]
            (instrument_ops) {Remove \enspace Replace \enspace Extract \enspace Double};
        \node[operations, right=1.15cm of drums]
            (drum_ops) {Remove \enspace Replace \enspace Extract};
        \node[operations, right=1.25cm of articulation]
            (articulation_ops) {Staccato \enspace Legato};
        \node[operations, right=1.25cm of dynamics]
            (dynamics_ops) {Crescendo \enspace Decrescendo};
        \node[operations, right=1.25cm of pitch]
            (pitch_ops) {Correction};

        \draw[edge] (root.east) -- (instrumentation.west);
        \draw[edge] (root.east) -- (articulation.west);
        \draw[edge] (root.east) -- (dynamics.west);
        \draw[edge] (root.east) -- (pitch.west);
        \draw[edge] (instrumentation.east) -- (instrument.west);
        \draw[edge] (instrumentation.east) -- (drums.west);
        \draw[edge] (instrument.east) -- (instrument_ops.west);
        \draw[edge] (drums.east) -- (drum_ops.west);
        \draw[edge] (articulation.east) -- (articulation_ops.west);
        \draw[edge] (dynamics.east) -- (dynamics_ops.west);
        \draw[edge] (pitch.east) -- (pitch_ops.west);
    \end{tikzpicture}%
    }
    \vspace{-1.5em}
    \caption{\textbf{SMEBench edit taxonomy.} Each of the twelve operations is sampled
    only when the excerpt contains an eligible target. Instrument-level
    dynamics and pitch operations may target one instrument or the complete
    mixture without changing the operation category.}
    \label{fig:edit_taxonomy}
\end{figure}
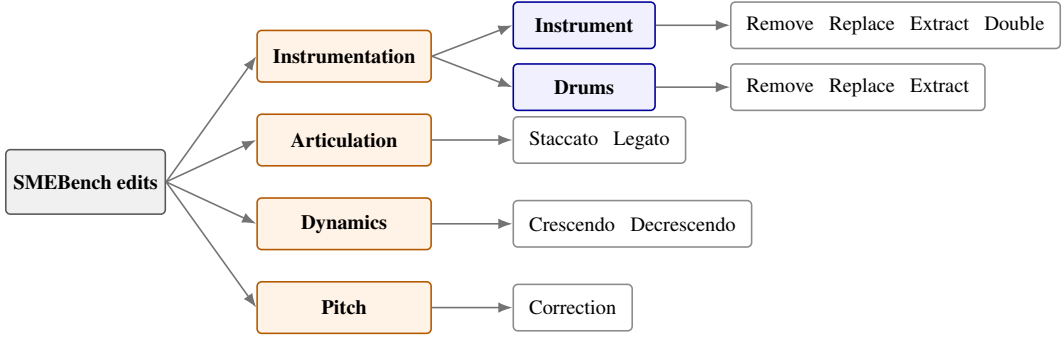

The released benchmark covers twelve operations across four edit families.
Instrumentation comprises removing, replacing, extracting, or doubling an
instrument and removing, replacing, or extracting drums. Articulation edits
shorten or lengthen notes to produce staccato and legato performances, while
dynamics edits apply a crescendo or decrescendo to a selected instrument or to
the complete mixture. The final operation corrects the tuning of a selected instrument or of the complete mixture: in the source, each note is detuned by an offset sampled between $-50$ and $+50$ cents relative to its nominal pitch, whereas the target keeps all notes on the twelve-tone equal-tempered grid referenced to $A_4 = 440$~Hz. Taken together, these edits cover
operations commonly performed in music composition, arrangement, and
production workflows. Appendix~\ref{app:edit_implementations} specifies the
exact transformations. Figure~\ref{fig:edit_taxonomy} summarizes the taxonomy.

Some edits are applicable to more than one instrument in an excerpt. In
these cases, we do not select uniformly among the eligible instruments.
Instead, we weight each candidate by its perceived usage, so that the target
more often corresponds to a prominent part of the excerpt while less prominent
instruments remain eligible. Usage combines the duration for which
an instrument is active with its MIDI velocity, following the relationship
between velocity and peak RMS amplitude described by
\citet{dannenberg2006midi_velocity}. Instruments that occupy more of the
excerpt and are played more strongly are therefore more likely to be selected,
without excluding quieter instruments altogether. The exact computation and
sampling distribution are given in Appendix~\ref{app:edit_sampling}.

After applying the selected intervention, we synthesize the source and target
MIDI with FluidSynth. We use five SoundFont libraries---\texttt{Arachno 1.0},
\texttt{FluidR3\_GM}, \texttt{General Montage}, \texttt{GeneralUser GS}, and \texttt{TimGM6mb}---which collectively
cover most General MIDI instruments represented in MIDICaps. A SoundFont is
sampled uniformly for each example and used for both the source and target, so
that the pair differs in the intended edit rather than in its renderer. Audio
is rendered in mono at 22,050 Hz with 32-bit floating-point precision.

We apply a final audio filter to remove ineffective and excessive
transformations. Lower and upper bounds on LPAPS and CLAP audio similarity
exclude pairs that are effectively unchanged and pairs whose source and target
differ too strongly. A minimum Audiobox Aesthetics score removes examples with
poor perceptual quality. These automatic checks are construction-time filters,
not evidence that the retained targets are perceptually correct; we evaluate
the quality, preservation, and instruction alignment of the retained examples
in Section~\ref{sec:experiments}. Appendix~\ref{app:rendering_filtering}
    reports the renderer configuration and filtering procedure.

The current taxonomy is one instantiation of the recipe rather than a fixed
limit of the construction. A new MIDI-expressible edit can be incorporated by
defining when it is applicable, how it transforms the symbolic performance,
and which metadata record the change. This separation between edit definition,
matched rendering, and annotation permits future extensions such as changes to
harmony, tempo, voicing, rhythm, or timbre while preserving the same benchmark
contract: a source and target related by a known intervention. Extensions
involving attributes that MIDI does not represent reliably, including vocals,
genre, or mixing and mastering, require complementary audio representations rather
than another symbolic edit alone.

\subsection{Captions}
\label{sec:captions}

The bottom branch of Figure~\ref{fig:smebench_pipeline} associates each audio
pair with four textual views. The source caption $c_{\mathrm{source}}$ and edit
caption $c_{\mathrm{edit}}$ describe the audio before and after the
intervention. They are constructed so that the edit caption changes only the
content affected by the operation. If the source caption describes ``a fast
piano melody with soft strings in the background,'' for example, an edit that
replaces the strings with brass yields ``a fast piano melody with soft brass
in the background.'' These captions describe their respective audio excerpts;
they are not phrased as transformations.
By contrast, the edit instruction $c_{\mathrm{instruct}}$ expresses the same intervention as a
user request. We generate it deterministically from manually written templates
for each edit type, using the selected instrument and operation recorded in the
edit metadata. The normalized caption $c_{\mathrm{norm}}$ instead gives a
standardized description of the resulting excerpt and explicitly lists its
instrumentation. We report text--audio scores under these different textual
conditions in Section~\ref{sec:experiments}, since editing systems and metrics
may consume either an imperative instruction or a description of the intended
output.

The descriptive captions are grounded in structured musical attributes rather
than generated directly from the complete MIDI file's original caption. A MIDICaps
caption may describe properties that do not hold within an arbitrary 10-second
crop, so we independently extract relevant descriptors from the source and target
MIDI and audio. These descriptors, the original MIDICaps caption, and the edit
metadata are provided to Qwen3-30B-A3B-Instruct-2507~\citep{qwen3technicalreport}, which
produces the source and edit captions and refines the normalized description.
An LLM-based judge then verifies that the instruction preserves the intended
operation and the target caption reflects the edit. It also checks that no
unrelated content is added or removed and that the source and target
descriptions remain consistent in detail. Entries that fail this check are discarded. The feature
extractors, prompts, and complete judge criteria appear in
Appendix~\ref{app:feature_xtract}. Because caption generation and filtering
use the same model family, we evaluate the retained captions independently in
Section~\ref{sec:experiments} and discuss the possibility of correlated errors
there.

\subsection{Benchmark Composition}
\label{sec:benchmark_composition}

Applying this recipe produces 11,160 paired edits in total. This corpus includes
a balanced test split of 3,600 examples, with 300 examples for each of the
twelve operations in Figure~\ref{fig:edit_taxonomy}. We release
the source and edited audio, their MIDI files, the instruction and descriptive
captions, the edit metadata, and MIDI masks representing the edited component
    and the remainder of the mixture.

We complement SMEBench with paired edits constructed from real multitrack recordings
from MoisesDB~\citep{pereira2023moisesdbdatasetsourceseparation}. After removing
tracks marked as containing bleed, we construct instrument-removal and
extraction edits by changing which stems are included in the mixture. For
doubling and swapping, we identify pairs of tracks within the same song that
play similar melodic or rhythmic material with different timbres, using
correlations between windowed chromagrams and onset-strength envelopes.
Dynamics edits are obtained by applying gain envelopes to selected stems.
Captioning and filtering follow the same principles as in the MIDI pipeline,
with MIDI-specific features omitted. This process yields a 416-example test set
from 107 tracks, which we denote SME-Real. The subset is
not intended to reproduce the coverage of the synthetic benchmark. MIDI and
virtual instruments are themselves widely used in music production, and
current systems remain separated from the paired-reference anchors even in this
controlled setting, before the additional complexity of real recordings. The
    real-audio complement tests whether our main findings persist when paired
    targets are constructed from recorded stems.

SMEBench and its real-audio complement offer different balances of control and
acoustic realism, but expose the same evaluation object: a source, a requested
change, and a paired target. Symbolic construction records which operation was
applied; it does not by itself establish that the rendered change is audible,
musically coherent, or accurately described. Section~\ref{sec:experiments}
therefore begins by validating the benchmark entries before using them to
diagnose metrics and compare editing systems.

\section{Experiments}
\label{sec:experiments}

Our experiments address three questions. We first test whether the paired
targets and captions are suitable for evaluation
(\Cref{sec:benchmark_validation}). We then use the known source and target
endpoints to examine how common audio-editing metrics behave
(\Cref{sec:metric_diagnosis}). Finally, we benchmark existing editing systems
and ask whether source-based, target-based, and text-based evaluation lead to
the same conclusions (\Cref{sec:system_benchmark}). Unless otherwise stated,
experiments use the balanced 3,600-example test split; we report results on the
416-example SME-Real test set separately.

\subsection{Validating Targets and Captions}
\label{sec:benchmark_validation}

Symbolic construction records which operation produced each target, but does
not establish that the rendered edit is perceptually convincing or correctly
described. We therefore conducted a listening study in which participants
heard 20 randomly selected source--target pairs together with the corresponding
edit instruction. On a five-point Likert scale, they rated the target's overall
quality, its alignment with the source, and its alignment with the edit
instruction. The latter measures whether the requested transformation is
audible in the target. For alignment with the source audio, participants were
explicitly asked whether the edited sample remained recognizable as the same
source apart from the requested modification, rather than whether it preserved
every mixture component. Participants also rated captions produced by our
pipeline: at each trial, either a source or an edit caption was randomly selected. The study involved 17 participants, of whom 9 completed all 20 trials,
yielding 236 trials in total.

Table~\ref{tab:user_study_summary} reports the mean ratings. Across 236 edit
ratings, targets received an average quality score of 4.23 and source-alignment
score of 4.65. Their mean instruction-alignment score was lower, at 3.99:
68.7\% of ratings were good or very good, while 13.2\% were bad or very bad.
This distinction matters for benchmark validity. The paired construction
preserves the intended relation at the symbolic level, but some rendered
transformations are less perceptible to listeners than others. Pipeline
captions received an average score of 3.70, with 71.2\% rated good or very
good. These results support using the retained targets and captions as
evaluation references, while also motivating an edit-wise analysis of their
reliability.

\begin{table}[H]
    \centering
    \caption{Human validation of the benchmark targets and captions. Ratings
    use a five-point scale. We report means over all available responses; the
    edit-wise breakdown appears in Appendix~\ref{app:user_study}.}
    \label{tab:user_study_summary}
    \small
    \begin{tabular}{lccc}
        \toprule
        Criterion & Mean & Responses & 95\% CI (participant bootstrap, n=17) \\
        \midrule
        Target quality & 4.23 & 236 & [4.06, 4.39]\\
        Alignment with source & 4.65 & 236 & [4.57, 4.73]\\
        Alignment with instruction & 3.99 & 236 & [3.72, 4.23]\\
        Pipeline-caption quality & 3.70 & 174 &  [3.46, 3.98]\\
        \bottomrule
    \end{tabular}
\end{table}

We report 95\% confidence intervals from a participant-clustered bootstrap
(5,000 resamples of whole participants, preserving within-participant
correlation across their trials). Three of the twelve operations
(\textsc{Legato}, \textsc{Remove\_Instrument}, and \textsc{Swap\_Drums}) do not
reliably exceed the midpoint of the instruction-following scale. We retain
these operations in the benchmark and report them separately, but exclude them
from the headline perceptual aggregate. Across the remaining nine operations
(164 trials, 17 participants), mean instruction following is 4.30
(95\% CI [4.03, 4.53]), source consistency is 4.69
([4.61, 4.77]), and overall quality is 4.43 ([4.29, 4.54]). The
edit-wise breakdown appears in
Appendix~\ref{app:user_study} (Table~\ref{tab:edit-quality-by-type}). Source consistency remains high ($\geq 4.3$) for the three flagged operations,
but listeners are less consistent in attributing the perceived result to the
requested instruction.

The caption comparison assesses the quality of retained captions; it does not
validate the rejection decisions made by the LLM filter. We therefore treat
correlated errors from using the same Qwen model family for caption generation
and filtering as a limitation rather than interpreting the study as validation
of the complete filtering procedure. We provide a short complementary analysis in
Appendix~\ref{app:llm_judge_analysis}.

\subsection{Diagnosing Music-Editing Metrics}
\label{sec:metric_diagnosis}

\begin{figure}[H]
    \centering
    \includegraphics[width=\textwidth]{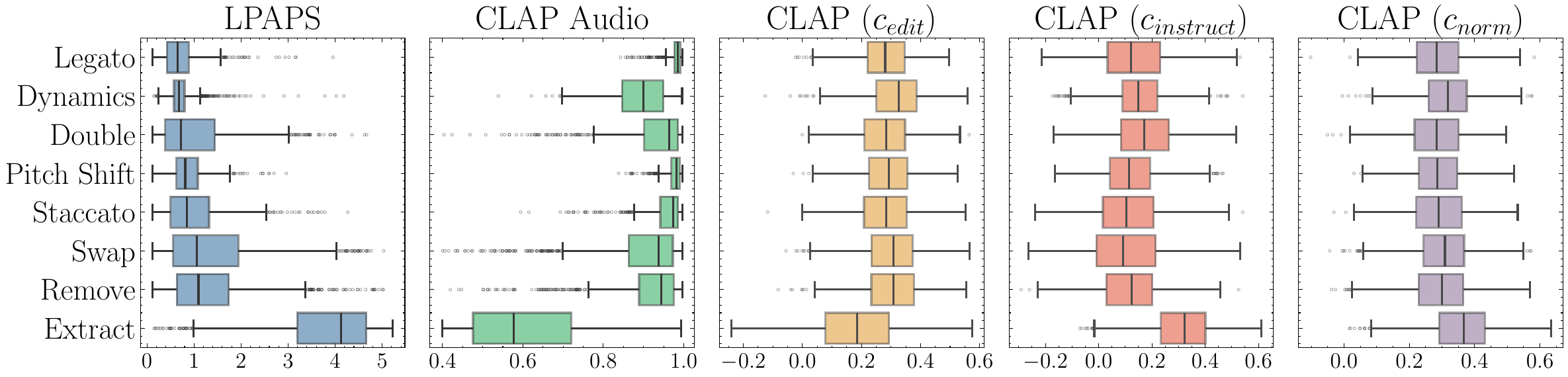}
    \caption{Metric distributions by edit group on the SMEBench test split.
    The twelve operations are displayed in eight groups: crescendo and
    decrescendo are grouped as Dynamics, while instrument and drum variants are
    grouped under Swap, Remove, and Extract. Boxes show the median and
    interquartile range; whiskers extend to 1.5 times the interquartile range.}
    \label{fig:metric_by_edit_current}
\end{figure}

Music editing requires both preservation and change, but commonly used metrics
observe these properties through different references. We distinguish three
evaluation axes. Source-reference metrics measure how much of the input is
preserved; target-reference metrics measure agreement with the known edited
outcome; and text--audio metrics measure compatibility with an instruction or
description. We instantiate the first two axes with CLAP-Audio similarity and
LPAPS distance, computed against either the source (Src) or paired target (GT),
and evaluate textual compatibility with CLAP-Text using the edit caption,
instruction, and normalized caption.

Identity and Ground Truth provide interpretable endpoints. Identity returns
the source unchanged and therefore represents perfect preservation without
edit execution. Ground Truth is the paired target produced by the known
intervention. A source-reference metric should prefer Identity, whereas a
target-reference metric should prefer Ground Truth. Text--audio similarity is
useful for editing only insofar as it can distinguish these endpoints when the
instruction describes their difference.

Table~\ref{tab:smebench_global_metrics_current} summarizes the global
source--target statistics. The source and target remain relatively close in the embedding spaces used by
current metrics. On the test set, their mean CLAP-Audio similarity is 0.873 and
their mean LPAPS distance is 1.513. The corresponding values on the real-audio
complement are 0.804 and 2.167, indicating a larger source--target separation
for the available real-audio edits. These aggregate values nevertheless hide
substantial variation across operations. Extraction edits alter the global
scene by retaining a single component, whereas articulation and pitch edits
change a more localized property; their distances should not be interpreted on
the same absolute scale.

\begin{table}[H]
    \centering
    \small
    \caption{Descriptive characterization of the source and paired target
    audio. These aggregate statistics quantify source--target separation but
    do not by themselves establish edit correctness.}
    \label{tab:smebench_global_metrics_current}
    \begin{tabular}{lcccc}
        \toprule
        & \multicolumn{2}{c}{Test set}
        & \multicolumn{2}{c}{Real-audio complement} \\
        Metric & Source & Target & Source & Target \\
        \midrule
        CLAP-Text (caption) $\uparrow$ & 0.249 & 0.278 & 0.134 & 0.123 \\
        CLAP-Text (instruction) $\uparrow$ & -- & 0.160 & -- & 0.125 \\
        CLAP-Text (normalized) $\uparrow$ & 0.291 & 0.308 & -- & -- \\
        Audiobox Aesthetics $\uparrow$ & 6.903 & 6.734 & 7.247 & 6.952 \\
        \midrule
        LPAPS (Src $\leftrightarrow$ GT) $\downarrow$
            & \multicolumn{2}{c}{1.513} & \multicolumn{2}{c}{2.167} \\
        CLAP-Audio (Src $\leftrightarrow$ GT) $\uparrow$
            & \multicolumn{2}{c}{0.873} & \multicolumn{2}{c}{0.804} \\
        \bottomrule
    \end{tabular}
\end{table}

Figure~\ref{fig:metric_by_edit_current} shows that metric values depend on the
requested operation. Extraction produces the largest LPAPS distances and
lowest CLAP-Audio similarities. Normalized-caption scores vary less across edit
groups than instruction and edited-caption scores. This stability is not by
itself evidence of a better editing metric; it may instead indicate lower
sensitivity to the requested edit. Additional salience-conditioned views appear
in Appendix~\ref{app:conditional_metrics}.

Together, these analyses illustrate the paper's central distinction: preservation,
target agreement, and textual compatibility are related but not
interchangeable. Without paired targets, a system can receive a favorable
source or text score even when the requested transformation is not the outcome
it approaches.

\subsection{Benchmarking Music-Editing Systems}
\label{sec:system_benchmark}
\begin{figure*}[t]
    \centering
    \includegraphics[width=\textwidth]{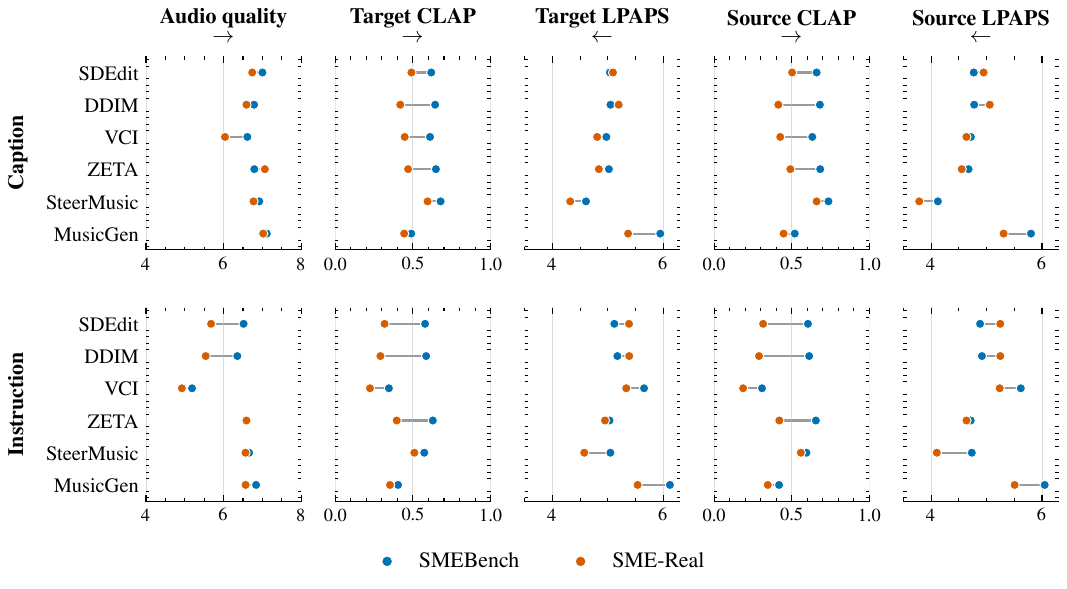}
    \vspace{-2.1em}
    \caption{
        System performance on SMEBench and SME-Real under edit-caption
        conditioning (top) and edit-instruction conditioning (bottom).
        Paired points represent the same editing pipeline on SMEBench
        (blue) and SME-Real (orange), with connecting lines showing the
        synthetic-to-real change. All panels report raw metric values;
        arrows indicate the preferred direction. Target and source metrics
        separately measure agreement with the intended edit and preservation
        of the source.
    }
    \label{fig:system_results_real_vs_synthetic}
\end{figure*}

We evaluate SDEdit~\citep{meng2022sdeditguidedimagesynthesis}, DDIM
inversion~\citep{song2022denoisingdiffusionimplicitmodels}, ZETA
\citep{manor2024zeroshotunsupervisedtextbasedaudio}, Virtual Consistency
Inversion (VCI)~\citep{cervera2025virtualconsistencyaudioediting}, and
SteerMusic~\citep{niu2025steermusicenhancedmusicalconsistency}. These methods
perform zero-shot editing with a pretrained diffusion model; we use the
\texttt{cvssp/audioldm2-music} AudioLDM2 checkpoint
\citep{liu2024audioldm2learningholistic} as their common generative backbone.
We additionally report MusicGen~\citep{copet2024simplecontrollablemusicgeneration}
as a text-conditioned generation baseline without source-preserving inversion.
We run the systems with either the imperative instruction
$c_{\mathrm{instruct}}$ or the edited-audio description $c_{\mathrm{edit}}$,
according to the input interface being evaluated.

We compare every output with both the source and target using CLAP-Audio and
LPAPS, and report CLAP-Text under the available textual conditions. We also
report Audiobox Aesthetics as a no-reference estimate of perceptual quality;
higher values are better.

Figure~\ref{fig:system_results_real_vs_synthetic} reports the principal
results, and Table~\ref{tab:ranking_reversals} makes the resulting ranking
reversals explicit. The radar comparison and remaining numerical results appear
in Appendix~\ref{app:full_system_results}.

\begin{table*}[t]
    \centering
    \small
    \setlength{\tabcolsep}{7pt}
    \begin{tabular}{llccc}
        \toprule
        Conditioning & System & CLAP-T $\uparrow$
            & CLAP-A GT $\uparrow$ & LPAPS GT $\downarrow$ \\
        \midrule
        Edit caption & SDEdit & 0.296 [0.293, 0.299]
            & 0.625 [0.621, 0.628] & 4.969 [4.951, 4.986] \\
        & DDIM & 0.283 [0.279, 0.286]
            & 0.652 [0.648, 0.655] & 4.981 [4.964, 4.998] \\
        & VCI & 0.299 [0.296, 0.303]
            & 0.614 [0.610, 0.618] & 4.908 [4.887, 4.929] \\
        & ZETA & 0.287 [0.284, 0.290]
            & 0.654 [0.651, 0.657] & 4.937 [4.918, 4.956] \\
        & SteerMusic & 0.269 [0.266, 0.272]
            & \textbf{0.691 [0.688, 0.694]}
            & \textbf{4.482 [4.458, 4.506]} \\
        & MusicGen & \textbf{0.350 [0.347, 0.353]}
            & 0.492 [0.489, 0.497] & 5.926 [5.914, 5.939] \\
        \midrule
        Instruction & SDEdit & 0.227 [0.223, 0.231]
            & 0.585 [0.580, 0.589] & 5.046 [5.028, 5.063] \\
        & DDIM & 0.203 [0.199, 0.207]
            & 0.596 [0.591, 0.601] & 5.094 [5.076, 5.111] \\
        & VCI & \textbf{0.266 [0.262, 0.269]}
            & 0.363 [0.357, 0.369] & 5.576 [5.555, 5.595] \\
        & ZETA & 0.199 [0.195, 0.203]
            & \textbf{0.633 [0.630, 0.637]}
            & \textbf{4.954 [4.936, 4.972]} \\
        & SteerMusic & 0.194 [0.191, 0.198]
            & 0.583 [0.579, 0.587] & 4.956 [4.934, 4.977] \\
        & MusicGen & 0.198 [0.194, 0.202]
            & 0.406 [0.402, 0.411] & 6.099 [6.086, 6.111] \\
        \bottomrule
    \end{tabular}
    \caption{System rankings on the SMEBench test split.
    The system preferred by CLAP-Text is not the system preferred by either
    paired-target audio metric. Values are means with 95\% bootstrap confidence
    intervals. Because some edits admit multiple valid realizations, target
    similarity measures agreement with the benchmark's known controlled
    realization rather than exhaustive edit correctness. Bold indicates the
    best observed mean within each conditioning mode and metric.}
    \label{tab:ranking_reversals}
\end{table*}

Under edit-caption conditioning, CLAP-Text favors MusicGen, whereas both
target-reference metrics favor SteerMusic. Under instruction conditioning,
CLAP-Text favors VCI, whereas both target-reference metrics favor ZETA. These
comparisons are diagnostic rather than exhaustive: some edit instructions
define a set of valid outputs, and the paired target represents one controlled
realization from that set. Nevertheless, the reversals show that broad textual
compatibility does not establish agreement with even a known valid realization
of the requested edit.

\begin{figure}[H]
    \centering
    \setlength{\tabcolsep}{2pt}
    \begin{tabular}{@{}cccc@{}}
        \textbf{Source} & \textbf{Paired target} & \textbf{SDEdit} & \textbf{DDIM} \\
        \includegraphics[width=0.235\textwidth]{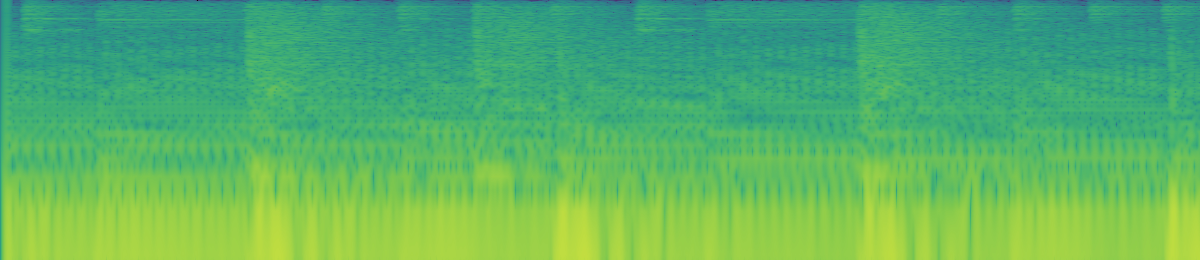} &
        \includegraphics[width=0.235\textwidth]{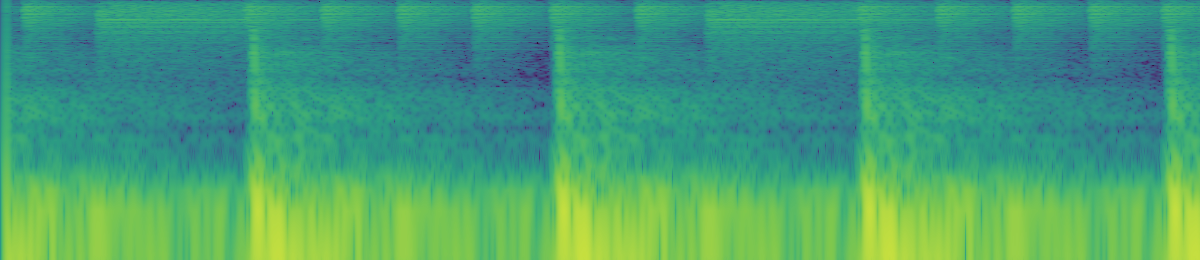} &
        \includegraphics[width=0.235\textwidth]{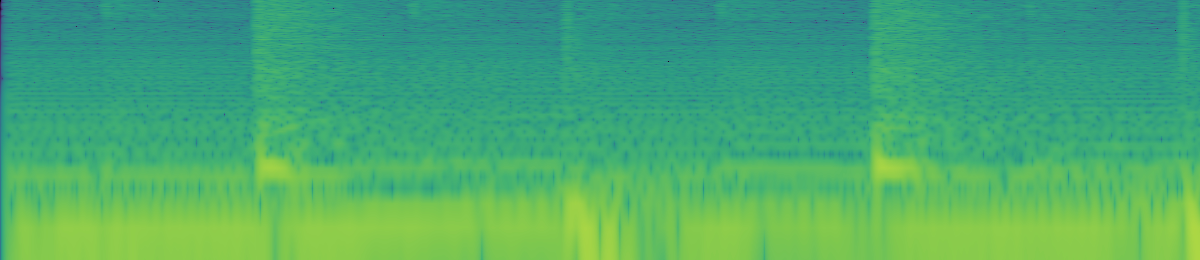} &
        \includegraphics[width=0.235\textwidth]{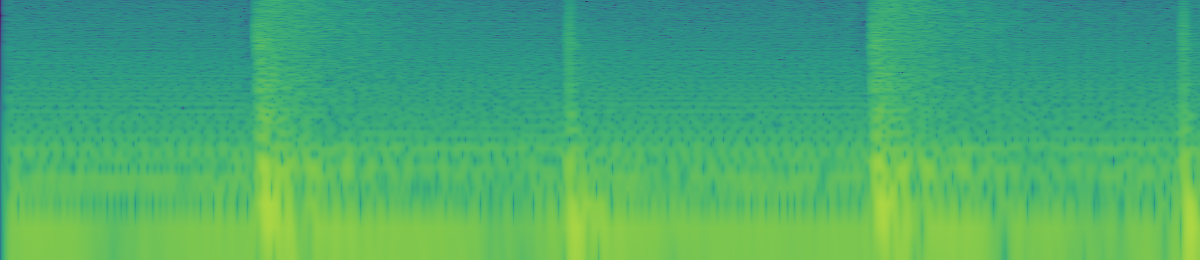} \\
        \textbf{VCI} & \textbf{ZETA} & \textbf{SteerMusic} & \textbf{MusicGen} \\
        \includegraphics[width=0.235\textwidth]{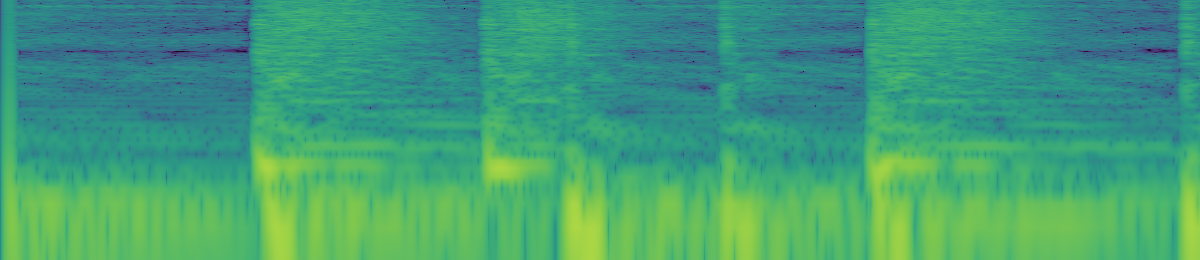} &
        \includegraphics[width=0.235\textwidth]{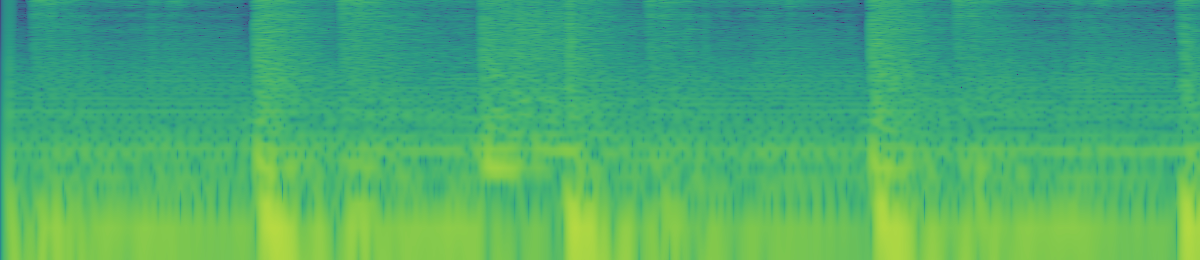} &
        \includegraphics[width=0.235\textwidth]{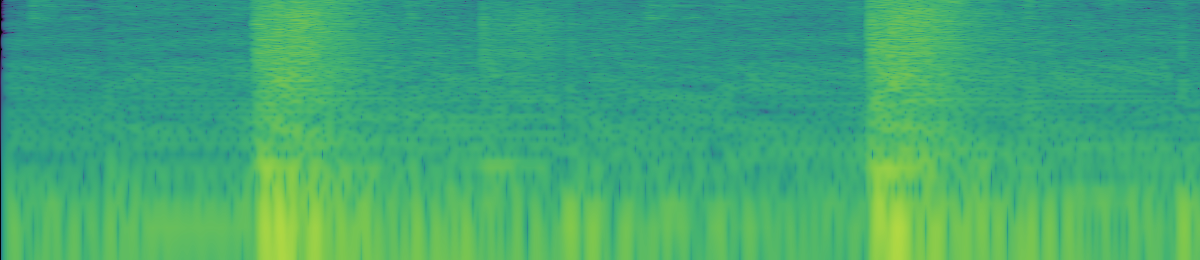} &
        \includegraphics[width=0.235\textwidth]{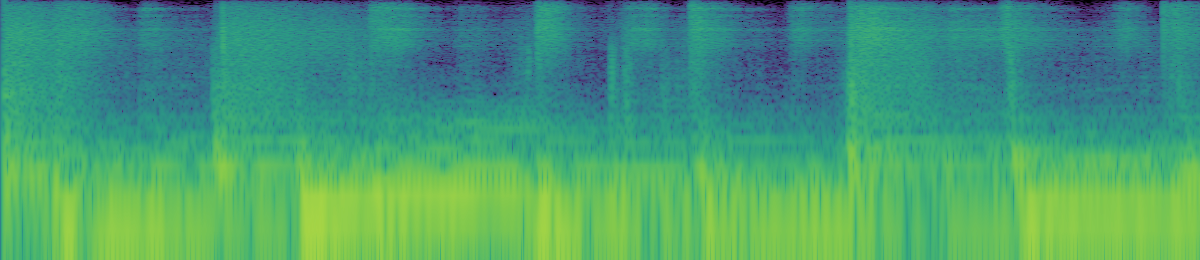}
    \end{tabular}
    \caption{Log-frequency spectrograms for one held-out SMEBench example.
    The paired target is one controlled valid realization; the remaining
    panels show the six system outputs. All panels use the same rendering
    configuration and show the same 2-second interval (4--6\,s).}
    \label{fig:qualitative_system_example}
\end{figure}

Figure~\ref{fig:qualitative_system_example} shows the corresponding evaluation
object for one held-out example. Differences in the outputs' time--frequency
structure illustrate that the methods do not converge to a single acoustic
solution; the example is illustrative rather than a correctness test.

\section{Conclusions and Limitations}
SMEBench provides paired source and target audio for controlled music edits.
Its symbolic intervention layer makes the requested change explicit and allows
preservation, target agreement, and text compatibility to be evaluated as
separate properties. The benchmark therefore complements realistic but
unpaired evaluation sets: it is designed not only to compare editors, but also
to reveal what their evaluation metrics reward. Our experiments show that
source-, target-, and text-based scores need not produce the same ranking,
supporting the use of paired targets as a diagnostic evaluation tool.

\paragraph{Limitations.}
The controlled rendering that makes the benchmark precise also creates a gap
from recorded music, which the smaller real-audio complement only partially
addresses. The current taxonomy covers twelve MIDI-expressible operations and
does not represent vocals, production effects, genre transfer, or other changes
that require richer audio representations. Some operations are less reliably
audible than others, and automatic audio and caption filtering can introduce
bias from the pretrained models on which it relies. We also note that music editing is not inherently bijective, so our synthesized edits represent one plausible solution among multiple valid alternatives.
Finally, no single metric captures preservation, edit execution, and perceptual quality; benchmark
results should therefore be interpreted across these axes rather than reduced
to one aggregate score.

\clearpage
\section*{Acknowledgments}
F. Paissan is partially funded by the Fonds de recherche du Qu\'ebec (FRQ)
through the grant ``Learning Interpretable Representations at Scale''
(\href{https://doi.org/10.69777/2009884}{doi:10.69777/2009884}) and by the
Institut intelligence et donn\'ees (IID) at Universit\'e Laval through its PhD
Admission Scholarship in Artificial Intelligence.
We acknowledge the support of the Natural Sciences and Engineering Research
Council of Canada (NSERC) and the Digital Research Alliance of Canada
(\url{https://alliancecan.ca}). Cem Subakan and Mirco Ravanelli are supported
by NSERC projects RGPIN-2023-05759 and RGPIN-2022-05088, respectively.
Mirco Ravanelli is also supported by Translated through the Imminent research grant. Micael Antunes is funded by the European Union through the Marie Skłodowska-Curie Actions (MSCA) Postdoctoral Fellowship GenCAST, Grant No. \href{https://doi.org/10.3030/101277782}{101277782}. Ricard Marxer benefited from the support of the French National Research Agency through the ANR-20-CE23-0012-01 (MIM) grant.

\section*{AI Use Statement}
In this work, we used generative AI tools to assist with drafting and editing
parts of the manuscript, improving readability and conciseness, typesetting,
formatting references, brainstorming, searching for relevant literature, and
implementing experimental methods. Generative models are also explicit
components of the reported methodology: Qwen3-30B-A3B-Instruct-2507 is used to
produce and assess grounded textual descriptions, and an independent Mistral
checkpoint is used to analyze judge agreement. We did not use generative AI to
formulate the benchmark's central claims, design its evaluation methodology,
or interpret the experimental results. The authors manually verified the
AI-assisted text and citations, inspected the generated captions and associated
quality-control procedure, and tested the experimental code. We take full
responsibility for the final manuscript, claims, experimental results, and
released artifacts.

\section*{Reproducibility Statement}
The benchmark construction and evaluation protocol are documented throughout
the paper and appendix. Section~\ref{sec:smebench} describes the paired-edit
recipe, edit taxonomy, caption construction, filtering procedure, and benchmark
composition. Appendices~\ref{app:edit_implementations},
\ref{app:edit_sampling}, and~\ref{app:rendering_filtering} specify the twelve
edit operations, their eligibility and target-sampling rules, the perceived-
usage computation, rendering configuration, SoundFonts, and construction-time
filtering procedure. Appendix~\ref{app:feature_xtract} provides the feature
extraction, captioning, and judge prompts. Section~\ref{sec:experiments}
defines the evaluation metrics, reference anchors, conditioning settings, and
editing systems, while Appendices~\ref{app:conditional_metrics} and
\ref{app:full_system_results} provide the detailed metric analyses and complete
system results. The released benchmark contains source and target audio, MIDI
files, instructions, captions, edit metadata, and component masks; the dataset
and accompanying generation code are made available through the
project release linked in the abstract.

{
\small

\bibliographystyle{iclr2027_conference}
\bibliography{refs}

}

\newpage
\appendix

\begin{center}
{\Large\bfseries Appendix}
\end{center}

\section{Audio Editing Pipelines in Related Work}
\label{app:editing_pipelines_rw}

\paragraph{Latent diffusion model-based audio editing.}
Several works in the literature apply generative models to audio editing, with diffusion models representing a prominent class of approaches. AUDIT~\citep{wang2023auditaudioeditingfollowing} and InstructME~\citep{han2023instructmeinstructionguidedmusic} use diffusion models to perform a closed set of editing instructions, such as \emph{Add Sound}, \emph{Remove Sound}, \emph{Replace Sound}, and \emph{Inpaint}. SmartDJ~\citep{lan2025guidingaudioeditingaudio} later proposed leveraging an LLM to determine a sequence of closed-set edits, improving the flexibility and quality of the resulting edits.

A second line of work explores training-free editing approaches that modify representations in the latent space of a latent diffusion model. For example, PPAE~\citep{xu2024promptguidedpreciseaudioediting} performs prompt-guided editing by manipulating latent representations, while~\citep{Paissan_2024} realizes edits through latent variable interpolation. SteerMusic~\citep{niu2025steermusicenhancedmusicalconsistency} instead steers the diffusion trajectory by optimizing the model score using a delta denoising score conditioned on the target edit prompt.

A third family of methods relies on diffusion model inversion to perform edits. These include ZETA~\citep{manor2024zeroshotunsupervisedtextbasedaudio}, SDEdit~\citep{meng2022sdeditguidedimagesynthesis}, and the virtual consistency approach of~\citep{cervera2025virtualconsistencyaudioediting}, which approximates inversion for audio editing.

\paragraph{Other editing approaches.}
Beyond diffusion-based editing, several works propose alternative interfaces and generative formulations. Recomposer~\citep{ellis2025recomposereventrollguidedgenerativeaudio} defines an autoregressive audio-editing pipeline conditioned on an event-activity interface. Other music generation and editing systems, including MusicLM~\citep{agostinelli2023musiclmgeneratingmusictext} and recent high-fidelity text-guided music editing approaches~\citep{lan2024highfidelitytextguidedmusic}, provide related directions for controllable music generation and transformation.

\subsection{Audio editing evaluation metrics}

 Audio editing involves modifying an input audio based on user-provided textual instructions. A central difficulty is maintaining a balance: the output should preserve relevant characteristics of the original audio, while faithfully reflecting the requested edits. Evaluation metrics must assess both fidelity to the input and alignment with the requested transformation. Currently, no single metric jointly captures both aspects.

\paragraph{Audio Fidelity Metrics}

Audio fidelity metrics often operate on learned audio embeddings and measure similarity at either the sample or distribution level. Sample-wise measures typically rely on cosine similarity in a shared embedding space, most commonly using contrastive language audio pretraining (CLAP)~\citep{elizalde2022claplearningaudioconcepts, wu2024largescalecontrastivelanguageaudiopretraining} representations, or on point-wise comparisons across intermediate layers~\citep{Paissan_2024}. Distributional metrics instead compare statistics of embedding distributions, often under a Gaussian assumption, using divergences such as Kullback–Leibler or the Fréchet Audio Distance (FAD)~\citep{kilgour2019frechetaudiodistancemetric}. While earlier approaches relied on CNN-based feature extractors, recent work~\citep{gui2024adaptingfrechetaudiodistance} has shown that CLAP-based embeddings yield FAD scores that correlate better with perceived acoustic and musical quality. Extensions of FAD beyond the Gaussian assumption have also been explored~\citep{chung2025kadfadeffectiveefficient}. In addition, no-reference metrics such as Meta AudioBox Aesthetics~\citep{tjandra2025metaaudioboxaestheticsunified} aim to predict perceptual quality by modeling multiple dimensions of audio perception, and have shown improved correlation with human mean opinion scores across speech, audio, and music.

\paragraph{Text-Consistency Metrics}
Text–audio consistency is measured using pretrained joint audio–text embedding models. The most common approach relies on CLAP embeddings. Alternative audio-language models, such as MuLAN \citep{huang2022mulanjointembeddingmusic}, can be used similarly; however, such approaches have not seen widespread adoption.

\section{Edit Implementations}
\label{app:edit_implementations}

\paragraph{Instrument and drum removal.}
Instrument removal deletes the selected instrument from the mixture. Drum
removal deletes all events assigned to MIDI channel 9, following the General
MIDI convention used by the source data. Instruments with percussive timbres
on other channels remain classified as instruments.

\paragraph{Instrument and drum replacement.}
Instrument replacement assigns the selected track to another program in the
General MIDI library. The replacement is sampled either from the same
instrument family or from a different family. Drum replacement analogously
maps a selected drum sound to another sound in the General MIDI percussion
map. Replacements are sampled subject to the same eligibility criteria used for
the other target-dependent operations.

\paragraph{Instrument and drum extraction.}
Instrument extraction retains the selected instrument and removes all other
tracks. Drum extraction removes the pitched instruments and retains the drum
events on channel 9. Because extraction can eliminate most of an excerpt, it
is admitted only when the retained content satisfies the eligibility criteria
in Appendix~\ref{app:edit_sampling}.

\paragraph{Instrument doubling.}
Instrument doubling creates a new track assigned to a randomly sampled
instrument, which plays the same notes as the selected source instrument.

\paragraph{Staccato.}
Notes are grouped into onset clusters whose members share one inter-onset
interval (IOI). The duration of each note is set to 40\% of its cluster's IOI
and clipped to the interval $[40,120]$ ms.

\paragraph{Legato.}
Each note is extended into the following onset cluster by 15\% of the IOI.
For monophonic selections, MIDI CC68, CC126, and CC127 messages are added to
activate legato and monophonic behavior and reinforce the glide between notes.

\paragraph{Crescendo and decrescendo.}
Dynamics edits are applied either to a selected instrument or to the complete
mixture. Editing MIDI velocity alone produced poor perceptual results, so the
gain envelope is applied to the synthesized audio. For a segment of duration
$T$, the envelope is
\begin{equation}
    g(t) = g_0 f(t/T),
\end{equation}
where $f(x)=x$ produces a crescendo and $f(x)=1-x$ produces a decrescendo.
This audio-domain implementation also illustrates how the recipe could be
extended to effects such as reverberation or delay, although such effects are
not part of the current release.

\paragraph{Pitch correction.}
To construct an out-of-tune source, the pitch of every note in the selected
instrument is independently perturbed by a value sampled uniformly from
$[-50,+50]$ cents. The unmodified, conventionally tuned MIDI is used as the
target. Pitch correction is applied either to one instrument or to all
instruments in the excerpt.

\section{Edit Eligibility and Sampling}
\label{app:edit_sampling}

For every 10-second MIDI excerpt, we determine the active instruments, drum
events, and their usage before constructing the set of admissible edits. An
edit is excluded when its target is absent or barely active, when the operation
would produce no perceptible change, or when following the instruction would
leave an empty or nearly empty target. For example, if an excerpt contains
only an acoustic grand piano, drum edits and instrument extraction are
discarded, leaving only edits that can meaningfully target the piano.

We estimate the perceived usage of an instrument from both note activity and
velocity. Let $\mathcal{N}_i$ be the set of notes active in frame $i$, let
$v_j \in [0,127]$ be the MIDI velocity of note $j$, and let $N$ be the number
of frames. The usage estimate is
\begin{equation}
    u_{\mathrm{percep}}
    = \frac{1}{N}\sum_{i=1}^{N}
      \max_{j \in \mathcal{N}_i}
      \left(\frac{v_j}{127}\right)^2.
\end{equation}
The quadratic velocity term follows the relationship between normalized MIDI
velocity and the square root of peak RMS amplitude described by
\citet{dannenberg2006midi_velocity}. Given $K$ instruments with non-zero
usage, the probability of selecting instrument $k$ is
\begin{equation}
    p(k) = \frac{u_{\mathrm{percep}}^{(k)}}
                 {\sum_{j=1}^{K}u_{\mathrm{percep}}^{(j)}}.
\end{equation}
This prioritizes salient instruments while retaining a non-zero probability
of editing less prominent ones.

\section{Audio Rendering and Filtering}
\label{app:rendering_filtering}

We synthesize source and edited MIDI files with FluidSynth using one of five
SoundFont libraries: Arachno 1.0, FluidR3\_GM, General Montage, GeneralUser GS,
and TimGM6mb. The same uniformly sampled SoundFont renders both members of a
pair. Audio is stored as a mono signal sampled at 22,050 Hz with 32-bit
floating-point precision. We verified on a large subset of edits that the
choice of SoundFont did not produce a significant shift in the distributions
of the general metrics used during construction.

We remove pairs with negligible changes by imposing a minimum LPAPS distance
and a maximum CLAP audio similarity. Conversely, a maximum LPAPS distance and
a minimum CLAP audio similarity remove edits that alter the source too
strongly. The thresholds are obtained by averaging the fifth-percentile values
from multiple large-scale construction runs (more than $20,000$ total samples). Finally, we require a minimum
mean Audiobox Aesthetics score to reject poor-quality audio.

The same percentile-based rule is applied consistently across construction
runs. Its purpose is to remove distributional extremes rather than to define a
universal perceptual boundary for a successful edit.

\section{Human-Evaluation Details}
\label{app:user_study}

Participants rated target quality, alignment with the source, alignment with
the instruction, and caption quality on a scale from 1 (very bad) to 5 (very
good). Table~\ref{tab:edit-quality-by-type} gives the edit-wise results used to
identify operations whose instruction-following ratings do not reliably clear
the midpoint of the scale.

\begin{table*}[t]
\centering
\small
\caption{Instruction-following, source consistency, and overall quality by
edit operation (mean and 95\% participant-clustered bootstrap confidence
interval). Operations marked $\dagger$ do not reliably clear the midpoint of
the instruction-following scale and are excluded only from the headline
perceptual aggregate.}
\label{tab:edit-quality-by-type}
\resizebox{\textwidth}{!}{%
\begin{tabular}{lrrccc}
\toprule
Edit operation & $n_{\text{trials}}$ & $n_{\text{part.}}$ & Instr.-following & Consistency & Overall \\
\midrule
Remove drums        & 11 & 11 & 4.27 [3.64, 4.82] & 4.82 [4.55, 5.00] & 4.91 [4.73, 5.00] \\
Crescendo            & 12 & 12 & 4.50 [4.00, 4.92] & 4.75 [4.50, 5.00] & 4.67 [4.42, 4.92] \\
Decrescendo          & 26 & 15 & 4.50 [4.11, 4.82] & 4.88 [4.70, 5.00] & 4.54 [4.23, 4.81] \\
Staccato             & 25 & 14 & 4.44 [4.04, 4.78] & 4.68 [4.46, 4.88] & 4.56 [4.28, 4.81] \\
Double instrument   & 24 & 14 & 4.42 [4.15, 4.70] & 4.83 [4.68, 4.96] & 4.54 [4.26, 4.79] \\
Extract drums       & 20 & 11 & 4.35 [4.00, 4.64] & 4.70 [4.53, 4.86] & 4.50 [4.19, 4.77] \\
Extract instrument  & 11 & 11 & 4.18 [3.18, 5.00] & 4.82 [4.55, 5.00] & 4.36 [3.82, 4.82] \\
Replace instrument  & 22 & 13 & 4.00 [3.61, 4.39] & 4.09 [3.87, 4.29] & 3.77 [3.57, 4.00] \\
Pitch correction    & 13 & 13 & 3.85 [3.23, 4.46] & 4.77 [4.38, 5.00] & 4.15 [3.77, 4.54] \\
Legato$\dagger$              & 15 & 15 & 3.33 [2.80, 3.87] & 4.60 [4.33, 4.87] & 3.93 [3.40, 4.40]\\
Remove instrument$\dagger$  & 46 & 15 & 3.27 [2.79, 3.73] & 4.60 [4.43, 4.77] & 3.78 [3.35, 4.18] \\
Replace drums$\dagger$         & 11 & 11 & 3.00 [2.40, 3.64] & 4.30 [3.90, 4.70] & 3.50 [2.90, 4.00] \\
\midrule
\textbf{Headline (excl. $\dagger$)} & \textbf{164} & \textbf{17} & \textbf{4.30 [4.03, 4.53]} & \textbf{4.69 [4.61, 4.77]} & \textbf{4.43 [4.29, 4.54]} \\
\bottomrule
\end{tabular}%
}
\end{table*}
\section{Metric Behavior by Edit and Target Salience}
\label{app:conditional_metrics}

\begin{figure}[t]
    \centering
    \includegraphics[width=0.48\textwidth]
        {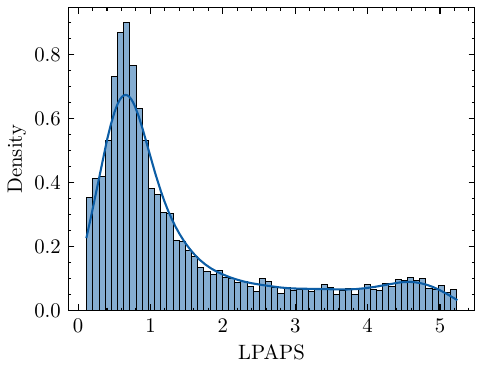}
    \includegraphics[width=0.48\textwidth]
        {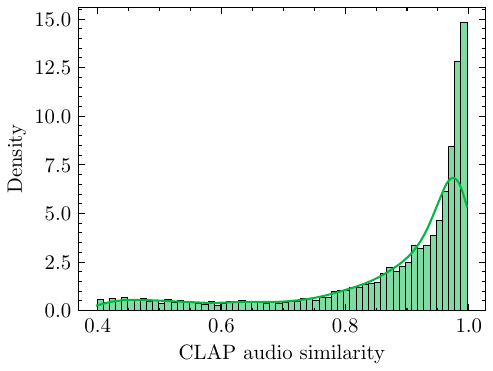}
    \caption{Source--target LPAPS distance and CLAP-Audio similarity
    distributions on the SMEBench test split ($n=11{,}160$).}
    \label{fig:metrics_global_original}
\end{figure}

\begin{figure*}[t]
    \centering
    \includegraphics[width=0.32\textwidth]
        {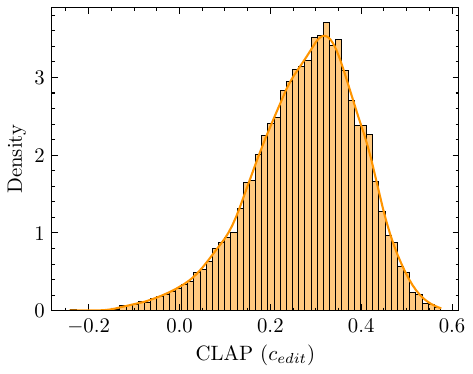}
    \includegraphics[width=0.32\textwidth]
        {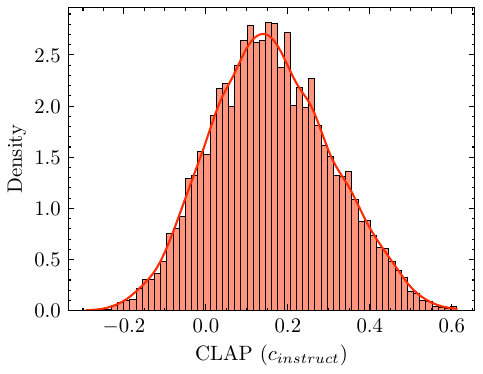}
    \includegraphics[width=0.32\textwidth]
        {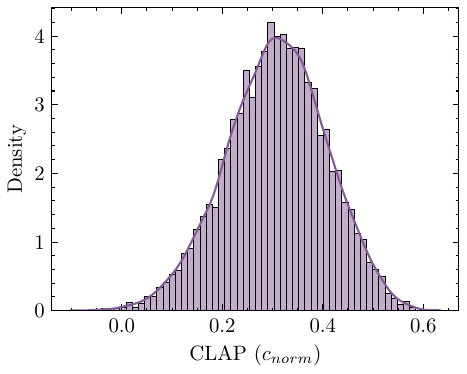}
    \caption{Global CLAP-Text distributions for the edit caption, instruction,
    and normalized caption on the SMEBench test split ($n=11{,}160$).}
    \label{fig:metrics_text_global_original}
\end{figure*}

The aggregate metric distributions conceal systematic differences among edit
types. In the current plots, extraction operations restructure the mixture by
retaining a single component and yield larger audio--audio distances than the other
instrumentation and drum operations. Figure~\ref{fig:metric_by_edit_current} reports
CLAP-Audio, LPAPS, and CLAP-Text distributions for all twelve operations,
grouped as described in its caption. The corresponding global audio and text distributions appear in
Figures~\ref{fig:metrics_global_original}
and~\ref{fig:metrics_text_global_original}.

\begin{figure*}[t]
    \centering
    \begin{minipage}[t]{0.46\textwidth}
        \centering
        \includegraphics[width=\linewidth]
            {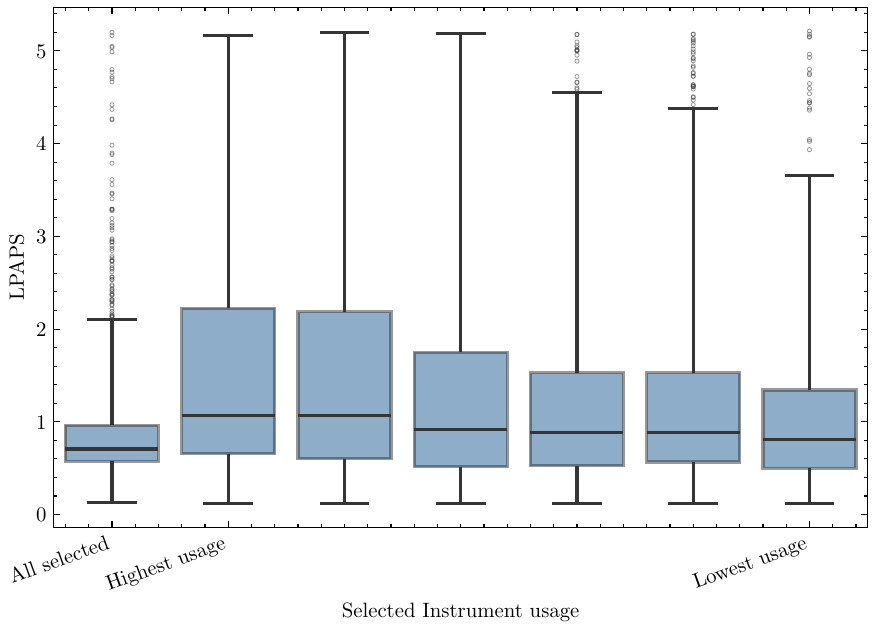}
        \par\smallskip
        \textbf{(a)} LPAPS by instrument usage
    \end{minipage}
    \hfill
    \begin{minipage}[t]{0.46\textwidth}
        \centering
        \includegraphics[width=\linewidth]
            {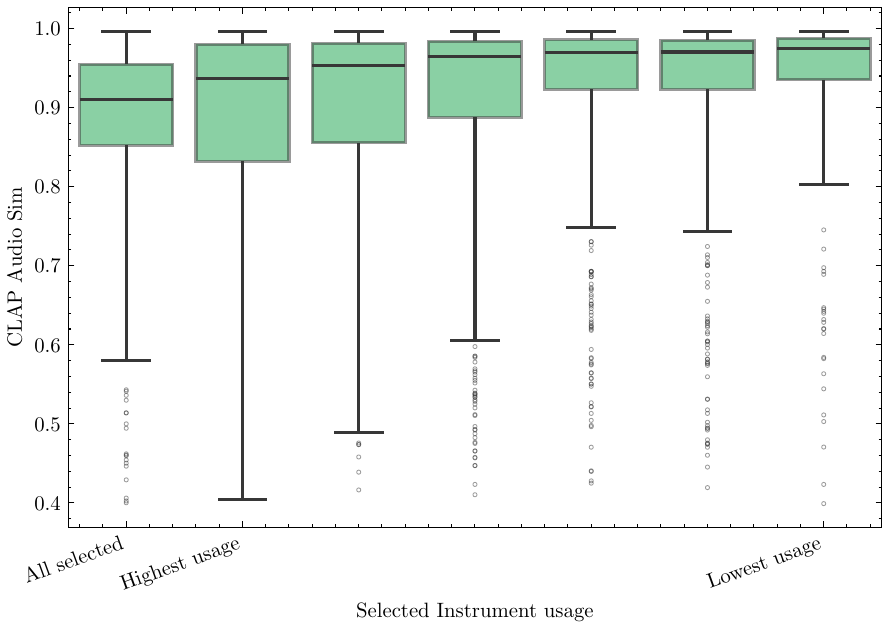}
        \par\smallskip
        \textbf{(b)} CLAP-Audio by instrument usage
    \end{minipage}
    \caption{Audio-similarity metrics by the within-excerpt perceived-usage
    rank of the edited instrument. Lower ranks correspond to targets with
    greater perceived usage.}
    \label{fig:metric_ranks_categories_original}
\end{figure*}

\begin{figure*}[t]
    \centering
    \includegraphics[width=0.32\textwidth]
        {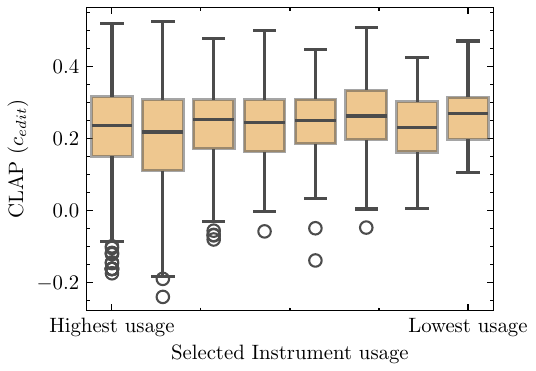}
    \includegraphics[width=0.32\textwidth]
        {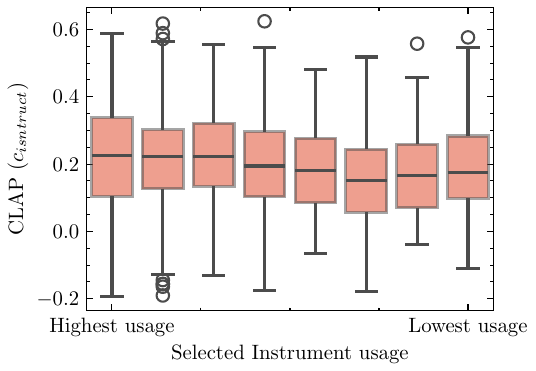}
    \includegraphics[width=0.32\textwidth]
        {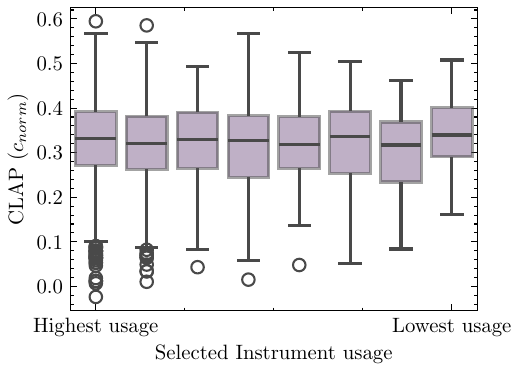}
    \caption{CLAP-Text distributions by the within-excerpt perceived-usage
    rank of the edited instrument. The analysis includes operations that target
    an individual instrument.}
    \label{fig:metric_ranks_text_original}
\end{figure*}

Metric values also depend on the perceived usage of the edited instrument.
Figure~\ref{fig:metric_ranks_categories_original} groups examples by the
within-excerpt perceived-usage rank used during target sampling; lower ranks
correspond to more prominent targets. This analysis tests whether a metric's
sensitivity is driven partly by the salience of the modified content rather
than the correctness of the operation itself. The analogous CLAP-Text
distributions appear in Figure~\ref{fig:metric_ranks_text_original}.

\section{Complete System Benchmark}
\label{app:full_system_results}

Tables~\ref{tab:system_audio_fidelity} and
\ref{tab:system_text_fidelity} report the complete results for SDEdit, DDIM,
VCI, ZETA, SteerMusic, and MusicGen under edit-caption and instruction
conditioning on the SMEBench and SME-Real test sets. Audio fidelity is
measured against both the source and paired target, while text fidelity is
measured against the edit caption, instruction, and normalized caption when
available.

\begin{figure}[t]
    \centering
    \includegraphics[width=\linewidth]
        {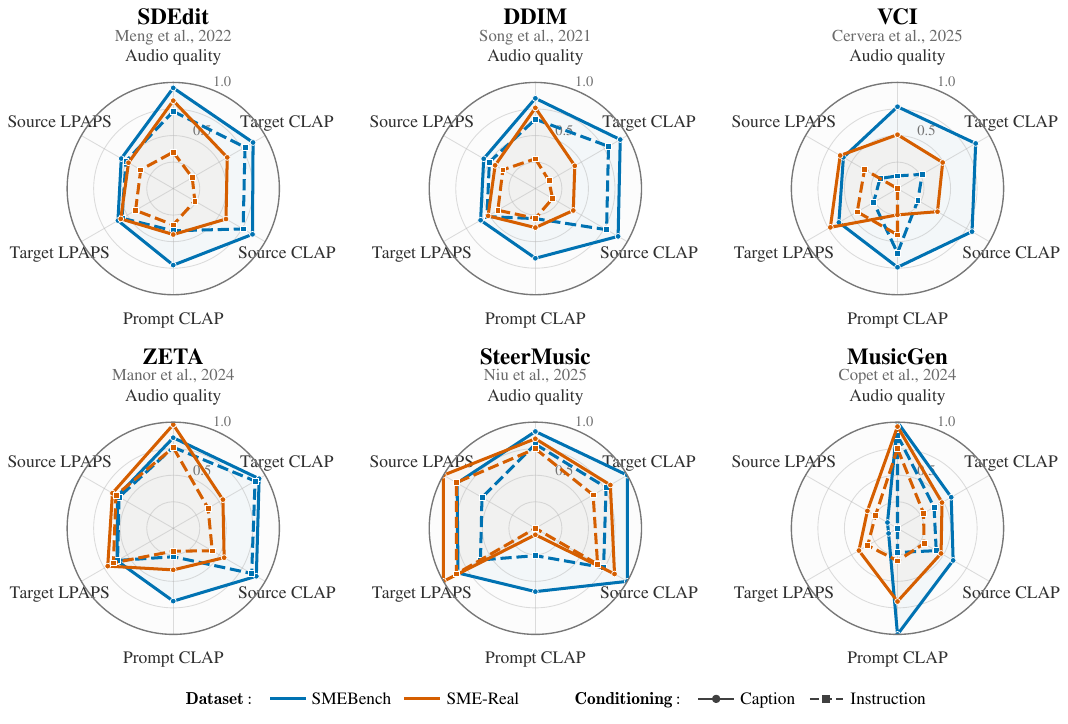}
    \caption{System-level comparison on SMEBench and SME-Real. Color
    distinguishes benchmarks, while line style and marker distinguish
    edit-caption from edit-instruction conditioning. Values are globally
    min--max normalized across the displayed systems, benchmarks, and
    conditions. LPAPS is reversed after normalization so that farther from the
    center consistently indicates better performance.}
    \label{fig:system_radar_appendix}
\end{figure}

Figure~\ref{fig:system_radar_appendix} provides a normalized, system-wise view
of the same results.

\begin{table*}[t]
    \centering
    \small
    \setlength{\tabcolsep}{3pt}
    \resizebox{\textwidth}{!}{%
    \begin{tabular}{lccccc}
        \toprule
        Method & Aesth. $\uparrow$ & CLAP-A GT $\uparrow$ & CLAP-A Src $\uparrow$
        & LPAPS GT $\downarrow$ & LPAPS Src $\downarrow$ \\
        \midrule
        \multicolumn{6}{l}{\textbf{Real-audio complement, edit-caption conditioning}} \\
        Identity & 7.247 [7.2010, 7.2926]  & 1.000 & 0.804 [0.7830, 0.8247]  & 2.167 [1.9977, 2.3384] & 0.000 \\
        Ground Truth & 6.947 [6.877, 7.016] & 0.804 [0.783, 0.824] & 1.000 & 0.000 & 2.167 [1.998, 2.338] \\
        SDEdit & 6.737 [6.675, 6.797] & 0.490 [0.474, 0.507] & 0.503 [0.488, 0.519] & 5.093 [4.990, 5.192] & 4.947 [4.842, 5.054] \\
        DDIM & 6.589 [6.497, 6.679] & 0.418 [0.404, 0.434] & 0.414 [0.399, 0.429] & 5.195 [5.091, 5.298] & 5.058 [4.951, 5.163] \\
        VCI & 6.036 [5.912, 6.146] & 0.447 [0.425, 0.468] & 0.426 [0.405, 0.448] & 4.806 [4.690, 4.918] & 4.638 [4.519, 4.756] \\
        ZETA & 7.062 [7.002, 7.122] & 0.469 [0.454, 0.483] & 0.492 [0.477, 0.506] & 4.837 [4.725, 4.945] & 4.552 [4.433, 4.669] \\
        SteerMusic & 6.770 [6.711, 6.829] & 0.594 [0.581, 0.608] & 0.661 [0.647, 0.674] & 4.321 [4.219, 4.427] & 3.786 [3.667, 3.900] \\
        MusicGen & 7.021 [6.984, 7.058] & 0.443 [0.430, 0.457] & 0.448 [0.434, 0.463] & 5.365 [5.272, 5.453] & 5.308 [5.218, 5.397] \\
        \midrule
        \multicolumn{6}{l}{\textbf{Real-audio complement, instruction conditioning}} \\
        SDEdit & 5.677 [5.573, 5.766] & 0.317 [0.301, 0.334] & 0.316 [0.299, 0.334] & 5.382 [5.279, 5.480] & 5.248 [5.141, 5.359] \\
        DDIM & 5.539 [5.405, 5.665] & 0.291 [0.272, 0.309] & 0.290 [0.272, 0.310] & 5.386 [5.280, 5.491] & 5.248 [5.141, 5.358] \\
        VCI & 4.924 [4.843, 5.012] & 0.223 [0.207, 0.240] & 0.187 [0.171, 0.204] & 5.332 [5.226, 5.439] & 5.237 [5.129, 5.344] \\
        ZETA & 6.589 [6.499, 6.677] & 0.395 [0.379, 0.411] & 0.421 [0.403, 0.439] & 4.948 [4.839, 5.052] & 4.638 [4.522, 4.758] \\
        SteerMusic & 6.564 [6.500, 6.629] & 0.509 [0.496, 0.524] & 0.559 [0.544, 0.575] & 4.575 [4.472, 4.673] & 4.103 [3.979, 4.216] \\
        MusicGen & 6.567 [6.499, 6.630] & 0.352 [0.338, 0.366] & 0.346 [0.332, 0.361] & 5.536 [5.445, 5.623] & 5.507 [5.416, 5.594] \\
        \midrule
        \multicolumn{6}{l}{\textbf{Test set, instruction conditioning}} \\
        Identity & -- & 0.873 [0.868, 0.878]  & 1.000 & 1.513 [1.470, 1.557] & 0.000 \\
        Ground Truth & 6.734 [6.716, 6.753] & 1.000 & 0.873 [0.868, 0.878] & 0.000 & 1.513 [1.470, 1.557] \\
        SDEdit & 6.517 [6.494, 6.538] & 0.585 [0.580, 0.589] & 0.608 [0.604, 0.613] & 5.046 [5.028, 5.063] & 4.859 [4.840, 4.876] \\
        DDIM & 6.354 [6.329, 6.379] & 0.596 [0.591, 0.601] & 0.619 [0.614, 0.624] & 5.094 [5.076, 5.111] & 4.893 [4.874, 4.912] \\
        VCI & 5.290 [5.263, 5.317] & 0.363 [0.357, 0.369] & 0.339 [0.332, 0.344] & 5.576 [5.555, 5.595] & 5.531 [5.508, 5.553] \\
        ZETA & 6.570 [6.550, 6.590] & 0.633 [0.630, 0.637] & 0.657 [0.653, 0.660] & 4.954 [4.936, 4.972] & 4.693 [4.675, 4.711] \\
        SteerMusic & 6.641 [6.623, 6.659] & 0.583 [0.579, 0.587] & 0.604 [0.600, 0.609] & 4.956 [4.934, 4.977] & 4.682 [4.658, 4.706] \\
        MusicGen & 6.822 [6.804, 6.839] & 0.406 [0.402, 0.411] & 0.419 [0.414, 0.423] & 6.099 [6.086, 6.111] & 6.042 [6.030, 6.055] \\
        \midrule
        \multicolumn{6}{l}{\textbf{Test set, edit-caption conditioning}} \\
        SDEdit & 6.983 [6.967, 6.999] & 0.625 [0.621, 0.628] & 0.660 [0.657, 0.663] & 4.969 [4.951, 4.986] & 4.762 [4.744, 4.779] \\
        DDIM & 6.766 [6.745, 6.785] & 0.652 [0.648, 0.655] & 0.683 [0.680, 0.687] & 4.981 [4.964, 4.998] & 4.759 [4.740, 4.776] \\
        VCI & 6.603 [6.580, 6.624] & 0.614 [0.610, 0.618] & 0.635 [0.630, 0.639] & 4.908 [4.887, 4.929] & 4.701 [4.681, 4.724] \\
        ZETA & 6.766 [6.749, 6.785] & 0.654 [0.651, 0.657] & 0.683 [0.680, 0.686] & 4.937 [4.918, 4.956] & 4.673 [4.653, 4.693] \\
        SteerMusic & 6.892 [6.875, 6.908] & 0.691 [0.688, 0.694] & 0.739 [0.736, 0.742] & 4.482 [4.458, 4.506] & 4.081 [4.057, 4.105] \\
        MusicGen & 7.111 [7.099, 7.122] & 0.492 [0.489, 0.497] & 0.517 [0.513, 0.521] & 5.926 [5.914, 5.939] & 5.815 [5.803, 5.827] \\
        \bottomrule
    \end{tabular}}
    \caption{System-level audio fidelity and quality with 95\% bootstrap confidence intervals. GT and Src denote comparisons against the paired target and source, respectively.}
    \label{tab:system_audio_fidelity}
\end{table*}

\begin{table*}[t]
    \centering
    \small
    \setlength{\tabcolsep}{3pt}
    \begin{tabular}{lccc}
        \toprule
        Method & CLAP-T Edit $\uparrow$ & CLAP-T Inst. $\uparrow$
        & CLAP-T Norm. $\uparrow$ \\
        \midrule
        \multicolumn{4}{l}{\textbf{Real-audio complement, edit-caption conditioning}} \\
        Ground Truth & 0.123 [0.114, 0.132]  & 0.125 [0.110, 0.140] & -- \\
        SDEdit & 0.235 [0.226, 0.245] & 0.150 [0.140, 0.161] & -- \\
        DDIM & 0.222 [0.212, 0.233] & 0.178 [0.167, 0.189] & -- \\
        VCI & 0.198 [0.185, 0.211] & 0.171 [0.159, 0.182] & -- \\
        ZETA & 0.227 [0.216, 0.239] & 0.165 [0.155, 0.175] & -- \\
        SteerMusic & 0.161 [0.149, 0.172] & 0.118 [0.105, 0.132] & -- \\
        MusicGen & 0.287 [0.277, 0.296] & 0.157 [0.148, 0.168] & -- \\
        \midrule
        \multicolumn{4}{l}{\textbf{Real-audio complement, instruction conditioning}} \\
        SDEdit & 0.102 [0.092, 0.113] & 0.217 [0.206, 0.228] & -- \\
        DDIM & 0.090 [0.076, 0.105] & 0.204 [0.193, 0.216] & -- \\
        VCI & 0.053 [0.040, 0.065] & 0.236 [0.223, 0.248] & -- \\
        ZETA & 0.155 [0.144, 0.167] & 0.192 [0.181, 0.202] & -- \\
        SteerMusic & 0.140 [0.129, 0.151] & 0.148 [0.136, 0.160] & -- \\
        MusicGen & 0.166 [0.155, 0.179] & 0.210 [0.198, 0.221] & -- \\
        \midrule
        \multicolumn{4}{l}{\textbf{Test set, instruction conditioning}} \\
        Ground Truth & 0.276 [0.273, 0.280] & 0.159 [0.155, 0.164]  & 0.308 [0.305, 0.312] \\
        SDEdit & 0.215 [0.211, 0.219] & 0.227 [0.223, 0.231] & 0.225 [0.221, 0.228] \\
        DDIM & 0.212 [0.208, 0.216] & 0.203 [0.199, 0.207] & 0.233 [0.230, 0.237] \\
        VCI & 0.106 [0.102, 0.111] & 0.266 [0.262, 0.269] & 0.162 [0.158, 0.166] \\
        ZETA & 0.242 [0.239, 0.246] & 0.199 [0.195, 0.203] & 0.273 [0.270, 0.276] \\
        SteerMusic & 0.218 [0.214, 0.222] & 0.194 [0.191, 0.198] & 0.235 [0.232, 0.238] \\
        MusicGen & 0.251 [0.247, 0.255] & 0.198 [0.194, 0.202] & 0.262 [0.259, 0.266] \\
        \midrule
        \multicolumn{4}{l}{\textbf{Test set, edit-caption conditioning}} \\
        SDEdit & 0.296 [0.293, 0.299] & 0.168 [0.164, 0.172] & 0.286 [0.282, 0.289] \\
        DDIM & 0.283 [0.279, 0.286] & 0.173 [0.169, 0.177] & 0.291 [0.288, 0.295] \\
        VCI & 0.299 [0.296, 0.303] & 0.175 [0.171, 0.179] & 0.299 [0.296, 0.302] \\
        ZETA & 0.287 [0.284, 0.290] & 0.174 [0.171, 0.178] & 0.311 [0.308, 0.314] \\
        SteerMusic & 0.269 [0.266, 0.272] & 0.166 [0.162, 0.170] & 0.271 [0.268, 0.274] \\
        MusicGen & 0.350 [0.347, 0.353] & 0.124 [0.120, 0.128] & 0.314 [0.311, 0.316] \\
        \bottomrule
    \end{tabular} %
    \caption{System-level text fidelity with 95\% bootstrap confidence intervals. Missing normalized captions for SME-Real are shown as dashes.}
    \label{tab:system_text_fidelity}
\end{table*}

\section{Sim-to-real perspective.}
\label{app:simtoreal}
The SMEBench test split is synthetic by design, but this is precisely what makes it useful as a diagnostic benchmark. Existing real-world evaluations, such as Table~1 of~\citet{cervera2025virtualconsistencyaudioediting}, report source-fidelity and text-alignment metrics on benchmarks like ZoME Bench and MedleyDB, where paired ground-truth edited audio is unavailable. These benchmarks are closer to deployment conditions, but their unpaired structure makes it difficult to determine whether a metric rewards the requested edit, source preservation, or general text-audio compatibility.

Our results provide a complementary sim-to-real perspective. In the SMEBench test split, the intended edit is known, so metrics can be examined in a controlled environment before being used on real, unpaired benchmarks. The fact that CLAP-Text scores are weakly sensitive to edit correctness, and that source- and target-reference metrics can favor different methods, suggests that real-world benchmark scores should be interpreted with care. Synthetic paired edits, therefore, do not replace real benchmarks; rather, they provide a controlled layer for characterizing metric behavior before applying the metrics in realistic settings.

\section{Feature Extraction and Captioning}
\label{app:feature_xtract}

Following \cite{bukey2026rethinkingmusiccaptioningmusic, Melechovsky2024}, we
caption our MIDI/audio files by constructing a structured metadata sheet before
feeding it to a language model. This prompting structure is designed to anchor
the LLM to the predicted metadata, reducing the risk of details unsupported by
the metadata profile. To build the fact sheet, we extract features from both
the MIDI and audio files.

We retrieve the General MIDI instruments and drums present in each excerpt,
together with its tempo and time signature. The five edit types without
instrumentation-based actions (pitch correction, crescendo, decrescendo,
staccato, and legato) are represented as tags attached to the affected
instrument, or to the whole excerpt when the edit is untargeted (\textit{e.g.},
a global crescendo). We also compute note density and overall dynamics, and
assign the corresponding tags when their thresholds are exceeded. We
complement the inherited whole-MIDI key with a symbolic, per-excerpt estimate
obtained through Krumhansl--Schmuckler key estimation on the excerpt's own
pitch-class distribution. When the two disagree and the local estimate is
sufficiently confident, the fact sheet reports the local estimate. If its
confidence falls below the threshold, the key is omitted rather than silently
repeating a song-level key that may not hold locally. We apply the same idea to
style. We compute an excerpt-level genre prediction by passing the audio
through the Discogs-trained MAEST model and its genre-classification head
(\textit{discogs-maest-10s-pw}; \citealp{alonso2023efficient}) using Essentia
\citep{alonso2020tensorflow}. Specifically, we extract embeddings from the
seventh transformer block of MAEST and pass them to the paired
\textit{genre\_discogs400} classification head, which produces probabilities
over 400 music-style labels. The prediction is reported alongside the
whole-song MIDICaps caption used as the style hint $c_\text{hint}$.

Both fact sheets, the edit name, the deterministic edit description $c_d$, and $c_\text{hint}$ are fed to the captioning LLM, instructed to (i) paraphrase $c_d$ into the imperative edit instruction $c_\text{paraphrase}$, (ii) reformulate $f_\text{source}$ into $c_\text{source}$, and (iii) reformulate $f_\text{edited}$ into $c_\text{edit}$. Crucially, $c_\text{edit}$ is constrained to reuse $c_\text{source}$ verbatim except for the minimal span reflecting the edited fact.

Finally, to bring variety and avoid the caption corpus over-representing a narrow set of surface forms (\textit{e.g.}, always reporting tempo as an exact BPM figure), fact sheets occasionally verbalize or use synonyms for selected attributes (\textit{e.g.}, \textit{andante} instead of \textit{92 BPM}), with the expression and wording choices drawn once per source/edited pair and shared between both sheets. Individual attributes are also randomly omitted from the fact sheets.

\section{Instructions to LLM}
\label{app:prompts}

We provide the instructions used by the captioning LLM (for caption generation
and normalized-caption rewriting) and the LLM-as-a-judge:

\textbf{Captioning LLM}

\begin{Verbatim}[breaklines,fontsize=\tiny]
You are given two fact sheets describing a 10-second MIDI excerpt before and after an edit, plus the edit itself.

- SOURCE_FACTS: {F0}
- EDITED_FACTS: {F1}
- Edit name: {En}
- Edit description: {E}
- Style hint (tone/mood only, may be unrelated to this excerpt): {S}
- Wording guidance for this caption (applies to SOURCE_DESCRIPTION's opening/phrasing; EDITED_DESCRIPTION inherits it via the copy-verbatim rule below): {W}

Tasks:
1. Knowing the edit name En, paraphrase the edit description (E') while preserving meaning.
    - Use imperative style, as if you were requesting the edit.
    - The paraphrase MUST reflect the operation type defined by En:
            * Swap_Instrument → substitution (original removed)
            * Double_Instrument → augmentation (original kept, new element added)
            * Remove_Instrument → deletion
            * Extract_Instrument → isolation (only the target element remains; all other elements are removed)
            * Swap_Drums → drums substitution
            * Remove_Drums → drums deletion
            * Extract_Drums → drums isolation
            * Crescendo, Decrescendo → dynamics changes
            * Legato, Staccato → articulation changes
            * Pitch_Shift → pitch fixing
            * CompositeEdit → sequential composition (apply multiple edits in a defined order; each edit modifies the result of the previous one)

2. Write SOURCE_DESCRIPTION: a natural, human-like caption of the source excerpt, using ONLY facts from SOURCE_FACTS.
    - You may borrow tone, mood or genre words from the style hint if they do not contradict SOURCE_FACTS.
    - Never state a fact (instrument, tempo, key, effect) that is not present in SOURCE_FACTS.
    - If the instrument/drum list is long and you shorten it for natural phrasing, keep the instrument or drum that the edit targets (named in the edit description, e.g. the one being swapped/removed/extracted/doubled, or bracket-tagged in the fact sheets) so its change is still traceable in EDITED_DESCRIPTION.

3. Write EDITED_DESCRIPTION: a natural caption, human-like of the edited excerpt, using ONLY facts from EDITED_FACTS.
    - EDITED_DESCRIPTION MUST reuse the exact same wording, phrasing, and sentence structure as SOURCE_DESCRIPTION, word for word, except for the minimal span of text needed to reflect the change described by EDITED_FACTS.
    - Do not paraphrase, reorder, or reword any part of SOURCE_DESCRIPTION that is unaffected by the edit -- copy it verbatim.
    - Avoid making it read like an aggregation of SOURCE_DESCRIPTION and the edit description.
    - Any instrument or drum that is PRESENT in EDITED_FACTS and directly involved in the edit -- named in the edit description, bracket-tagged (e.g. [staccato], [legato], [crescendo], [decrescendo], [out-of-tune]) in EDITED_FACTS, newly appearing (Swap_Instrument/Swap_Drums's replacement, Double_Instrument/Double_Drums's addition), or the surviving target of Extract_Instrument/Extract_Drums -- MUST be explicitly named in EDITED_DESCRIPTION, with its resulting state described in natural language (never using the bracket notation itself, and never as a comparison to how it was before -- state it as a plain current fact). It must never be silently dropped or folded into a generic "instruments" mention.
    - Any instrument or drum that DISAPPEARS between SOURCE_FACTS and EDITED_FACTS (Remove_Instrument/Remove_Drums's target, or the non-target elements removed by Extract_Instrument/Extract_Drums) must be the opposite: simply omit it from EDITED_DESCRIPTION entirely, restructuring the copied sentence only as much as needed to read naturally without it. Never narrate the removal itself (e.g. never write "the piano has been removed", "no longer present", "is now missing") -- the missing mention is itself how the change is conveyed, exactly as if that instrument had never been part of the excerpt.

Constraints:
- Never invent instruments, tempo, key, note density, dynamics, articulation, or effects not present in the given fact sheet.
- If either instrument/drum list is long enough that you need to trim it for readability, always drop instruments/drums that are unaffected by the edit first -- never drop the one the edit targets.
- You may naturally mention this is a short/10-second excerpt.
- EDITED_DESCRIPTION must read as a fresh, standalone description of the edited excerpt's own audio -- never as a diff against the source. Avoid comparison/before-after language (e.g. "instead of", "now", "no longer", "used to", "previously", "has been", "replaced", "added", "removed"); describe whatever is or isn't present as plain current fact.
- Keep factual consistency and neutral, descriptive language.
- Keep the phrasing natural.
- Keep SOURCE_DESCRIPTION and EDITED_DESCRIPTION at the same level of detail.
- In the case of multiple edit descriptions, address each sequentially.
- Never mention or imply uncertainty, confidence, estimation, or prediction (e.g. avoid words like "estimated", "predicted", "may be", "possibly", "uncertain", "suggests"). Every fact sheet fact is given to you already resolved -- state it plainly, as simply true.
- Never state an exact note-density or notes-per-second number. Use only the qualitative density word from the fact sheet (e.g. "sparse", "dense"), if one is given.
- Genre/style cue words from the fact sheet or style hint may be reused directly, including verbatim, when they fit naturally. But do not stack more than one or two of them together with "and" like a list of tags -- weave at most one or two into the mood/atmosphere phrasing.
- The fact sheet's genre cue is formatted as "genre: X, subgenre: Y" -- that is field labeling for your own reading, not caption text. Never output the literal words "genre:" or "subgenre:"; only use the genre/subgenre names themselves (X and Y).

### Examples
The examples below cover different edit types and are deliberately written with different sentence structures. Use them only to learn the minimal-diff rule (copy SOURCE_DESCRIPTION into EDITED_DESCRIPTION except for the smallest span needed) -- do not let any one example's shape become the default template you fall back to for every caption.

### Example (per-instrument Staccato edit):
SOURCE_DESCRIPTION: A moody, introspective short excerpt with a moderate trumpet and clarinet duet in F major at 96 BPM, over a steady, soft rhythm section.
EDITED_DESCRIPTION: A moody, introspective short excerpt with a moderate trumpet and short, staccato clarinet duet in F major at 96 BPM, over a steady, soft rhythm section.
(Only the clarinet's articulation is called out -- the trumpet and everything else stays exactly as written in SOURCE_DESCRIPTION.)

### Example (global Crescendo edit):
SOURCE_DESCRIPTION: A 10-second excerpt evoking a dark and relaxing cinematic atmosphere featuring dense, legato-like sitar melodies in G major, played at a slow 70 BPM in 4/4 time with moderate dynamics.
EDITED_DESCRIPTION: A 10-second excerpt evoking a dark and relaxing cinematic atmosphere featuring dense, legato-like sitar melodies in G major, played at a slow 70 BPM in 4/4 time, with a gradual increase in volume.
(Notice EDITED_DESCRIPTION repeats SOURCE_DESCRIPTION word-for-word, only replacing "with moderate dynamics" with the crescendo phrase "with a gradual increase in volume" -- nothing else changed. Also notice this is just one example's opening, not the template -- see the Wording guidance field for how to open this particular caption.)

### Example (Swap_Instrument edit):
SOURCE_DESCRIPTION: Sparse, gentle 10-second music built around an acoustic guitar melody over a steady kick-and-snare groove, in D minor at 84 BPM, with a wistful, low-key feel.
EDITED_DESCRIPTION: Sparse, gentle 10-second music built around an electric piano melody over a steady kick-and-snare groove, in D minor at 84 BPM, with a wistful, low-key feel.
(Only "acoustic guitar" is swapped for "electric piano" -- the original instrument is fully replaced, not added alongside, matching Swap_Instrument's substitution semantics.)

### Example (Decrescendo edit, note the different sentence structure and opening compared to the Crescendo example above):
SOURCE_DESCRIPTION: Played at a bright 128 BPM, this house music excerpt layers a punchy synth bassline with a crisp hi-hat groove, sitting at a steady, energetic volume throughout.
EDITED_DESCRIPTION: Played at a bright 128 BPM, this house music excerpt layers a punchy synth bassline with a crisp hi-hat groove, fading down in volume as it goes.
(Same facts and instruments as a decrescendo edit would need, but written with a different opening and rhythm than the Crescendo example -- this variety is expected, not the exception.)

### Example (Remove_Instrument edit -- notice there is NO removal narration):
SOURCE_DESCRIPTION: A relaxed passage led by a warm piano melody over a soft acoustic guitar strum, in A minor at 90 BPM, with gentle, sparse dynamics.
EDITED_DESCRIPTION: A relaxed passage led by a soft acoustic guitar strum, in A minor at 90 BPM, with gentle, sparse dynamics.
(The piano is simply gone -- EDITED_DESCRIPTION never says it was "removed" or "no longer present"; the clause naming it is dropped and the sentence is restructured only enough to still read naturally, everything else copied verbatim.)

Output:
EDIT_PARAPHRASE: ...
SOURCE_DESCRIPTION: ...
EDITED_DESCRIPTION: ...
\end{Verbatim}

\textbf{Normalized Caption refactoring}

\begin{Verbatim}[breaklines,fontsize=\tiny]
You are an expert in music production and MIDI file analysis. Your task is to rewrite the following MIDI instrument caption to make it more fluid and human-readable, while preserving all the instrument names and logically grouping similar instruments.

### Rules:
1. **Preserve all instrument names** in the output. Do not omit any.
2. **Group similar instruments** to make the caption more fluid. For example:
- "high bongo, low bongo" → "high and low bongos"
- "crash cymbal 1, crash cymbal 2" → "crash cymbals"
- "closed high hat, open high hat, pedal high hat" → "closed, open, and pedal high hats"
3. **Keep the structure** of the caption: start with the duration, followed by "music featuring", then list the instrument categories and their grouped instruments, and finally describe the rhythmic beat.
4. **Use natural phrasing** for lists (e.g., "A, B, and C" instead of "A, B, C").
5. **Do not add or remove any instruments**—only group and rephrase for readability.
6. Remove any number.
7. Maintain the original meaning while improving flow.
8. **Preserve any bracketed edit tags** exactly as given (e.g. "[out-of-tune]", "[staccato]", "[decrescendo]") immediately before the instrument name they describe. Never drop, reword, translate, or move them away from their instrument.
9. **Preserve any leading edit-descriptor word** appearing right after the duration and before "music" (e.g. "10s decrescendo music"). Keep that exact word in the rewritten sentence, worded naturally (e.g. "A 10s decrescendo piece of music featuring...").

### Example:
**Input:**
Music featuring bass timbres (plucked electric bass, slap bass 1); brass timbres (brass section, trumpet); ensemble timbres (choir "aah", orchestral hit, synth strings 2); guitar timbres (acoustic steel guitar); organ timbres (hammond organ); sound effects (sfx) (woodblock). The rhythmic beat is composed of hats (cabasa, closed high hat, cowbell, crash cymbal 1, crash cymbal 2, high bongo, low bongo, mute high conga, open high conga, open high hat, pedal high hat, short guiro, tambourine); kicks (high floor tom, low tom); snares (acoustic snare, hand clap, high mid tom, low mid tom, side stick).

**Output:**
A piece of music featuring bass timbres (plucked electric bass, slap bass); brass timbres (brass section, trumpet); ensemble timbres (choir "aah", orchestral hit, synth strings); guitar timbres (acoustic steel guitar); organ timbres (Hammond organ); and sound effects (woodblock). The rhythmic beat is composed of hats (closed, open, and pedal high hats, cabasa, cowbell, crash cymbals, high and low bongos, mute and open high congas, short guiro, tambourine); kicks (high floor tom, low tom); and snares (acoustic snare, hand clap, high and low mid toms, side stick).

### Example (with a deterministic edit tag):
**Input:**
10s decrescendo music featuring bass timbres (plucked electric bass); brass timbres ([out-of-tune] trumpet). The rhythmic beat is composed of hats (cowbell).

**Output:**
A 10s decrescendo piece of music featuring bass timbres (plucked electric bass) and brass timbres (out-of-tune trumpet). The rhythmic beat is composed of hats (cowbell).

\end{Verbatim}

\textbf{LLM-as-a-Judge}

\begin{Verbatim}[breaklines,fontsize=\tiny]
Role: You are a strict, unbiased evaluator. Your task is to verify whether the provided inputs satisfy all specified requirements.

Context: D0 and D1 are both captions of the same fixed excerpt -- D1 is meant to reuse D0's wording verbatim except for the minimal span needed to reflect edit E. Treat a difference in D1 as a genuine violation only when it changes what is being claimed about the excerpt (a new/missing instrument, sound source, or event; a materially different amount or specificity of detail). Treat a difference that merely rewords, reorders, or resynonymizes a claim D0 already made -- including mood/genre/style adjectives -- as compliant, not as new or reduced information.

Inputs:
Original description (D0)
Edited description (D1)
Original edit instruction (E)
Paraphrased edit instruction (E_prime)

Tasks:
1. List all differences between D0 and D1 concisely.
2. Evaluate each of the following criterion independently using only the following binary format:
    [CRITERION_NAME]: YES (if satisfied)
    [CRITERION_NAME]: NO (if not satisfied)

PARAPHRASE_CORRECT: E_prime preserves the meaning of E and uses imperative/request style.
EDIT_INSTRUCTION: E_prime is a coherent edit instruction for a text based audio editing pipeline.
EDIT_APPLIED: The requested edit (E, E_prime) is correctly reflected in the description D1.
NO_HALLUCINATION: D1 introduces no new sound event, instrument, or entity that D0 and E give no basis for. A differently-worded restatement of something D0 already describes (synonym, added mood/genre adjective, rephrasing) is not itself a new entity.
MINIMAL_CHANGE: every part of D1 unrelated to the required edit expresses the same claims as the corresponding part of D0, even if reworded. Judge the underlying claims, not the exact wording.
CONSISTENCY: D1 is coherent and factually consistent.
LEVEL_OF_DETAIL: D1 states the same number and specificity of facts as D0 (aside from the fact changed by the edit). Judge how much is claimed, not how it is phrased.

Constraints:
- Evaluate each criterion independently.
- Binary output : Output only one of these two options per criterion: "[CRITERION_NAME]: YES" or "[CRITERION_NAME]: NO"
- For each criterion, output only "[CRITERION_NAME]: YES" if the criterion is fully satisfied.
- For each criterion, output only "[CRITERION_NAME]: NO" if the criterion is not satisfied or cannot be confirmed.
- Do not include explanations, justifications, or additional text.
- The evaluation is only for judging the provided inputs.


Output:
{
"differences": "...",
"paraphrase_correct": "...",
"edit_instruction": "...",
"edit_applied": "...",
"no_hallucination": "...",
"minimal_change": "...",
"consistency": "...",
"level_of_detail": "...",
}
'''

input : "Base audio description: {source_caption} \nEdited description: {edited_caption} \nOriginal edit: {base_instruction} \nParaphrased edit: {edit_instruction}"
\end{Verbatim}

\section{LLM-as-a-judge Qwen checkpoint analysis}
\label{app:llm_judge_analysis}

To assess whether our choice to use the same Qwen checkpoint for both captioning and filtering introduces too much bias, we judge 8,461 generated captions with our original Qwen judge and an independent Mistral Instruct checkpoint Mistral-Small-24B-Instruct-2501~\citep{mistral2025}. Both judges used the exact same rubric and per-criterion prompts. We report in Table~\ref{tab:judge-agreement} raw agreement and Cohen's $\kappa$, which accounts for chance agreement under the judges' marginal acceptance rates.

\begin{table}[t]
\centering
\caption{Agreement between the original Qwen (\emph{ref}) judge and Mistral (\emph{mistral}) judge on 8,461 generated captions. Entries in the agreement matrix show the number of captions, with percentages in parentheses.}
\label{tab:judge-agreement}
\begin{tabular}{lcc}
\hline
 & \textbf{Mistral accepts} & \textbf{Mistral rejects} \\
\hline
\textbf{Ref accepts} 
& 3,228 (38.2\%) 
& 492 (5.8\%) \\
\textbf{Ref rejects} 
& 2,066 (24.4\%) 
& 2,675 (31.6\%) \\
\hline
\end{tabular}

\vspace{2mm}

\begin{tabular}{lc}
\hline
\textbf{Summary statistic} & \textbf{Value} \\
\hline
Overall agreement & 69.8\% \\
Ref acceptance rate & 44.0\% \\
Mistral acceptance rate & 62.6\% \\
Cohen's $\kappa$ & 0.41 \\
McNemar's test & $p < 10^{-16}$ \\
\hline
\end{tabular}
\end{table}

The judges reach the same decision on 69.8\% of captions
($\kappa=0.41$). Mistral accepts more items overall, including 80\% of
discordant pairs, which does not support the hypothesis that the Qwen judge
simply favors outputs from its own checkpoint family. Agreement varies across
criteria and edit types, with the largest differences for
\textit{minimal\_changes}, \textit{no\_hallucination}, and
instrument-isolation edits.

The independent judge therefore provides a complementary check on the retained
captions while also showing that judge choice mainly changes the strictness of
the filter. Our construction uses the more conservative Qwen decisions: in this
setting, rejecting some valid captions is preferable to retaining descriptions
that introduce unsupported content or fail to isolate the requested change.
We nevertheless treat judge-dependent acceptance as a limitation of automatic
caption filtering.

\section{Licensing and ethics}

\textbf{License}

SME-Bench is licensed under CC BY-NC-SA 4.0. SME-Bench incorporates and is derived from multiple third-party resources. Each resource remains subject to its original license. Users of this dataset must comply with all applicable licenses listed below. Here are the following third-party resources. GeneralUser GS and FluidR3-GM are distributed under CC BY 4.0; TimGM6mb is in the public domain; Arachno SoundFont is distributed under CC BY-NC-SA 4.0; and General Montage is distributed under CC BY 3.0. The MIDI datasets used in our experiments are MidiCaps (CC BY-SA 4.0) and MoisesDB (CC BY-NC-SA 4.0). We use Qwen3-30B-A3B-Instruct-2507, distributed under the Apache License 2.0, for captioning and judging. Appropriate attribution and license requirements are preserved in accordance with the respective licenses and can be found at \url{https://huggingface.co/datasets/anonymous-account-submission/SME-Bench}.

\textbf{Distributional Gaps and Underrepresented Populations}

 The MIDI data is predominantly derived from Western popular music traditions, including mainly classical, orchestral, electronic, rock and pop genres. As a result, the dataset underrepresents non-Western musical styles, instruments, and cultural traditions. The distribution of instruments, both in the original MIDI files and the synthesized versions, is heavily biased toward Western musical conventions. Additionally, the soundfonts used for synthesis are optimized for Western instrument timbres, further limiting the dataset's cultural and stylistic diversity.

\textbf{Intended use case}

This dataset is designed to train and evaluate neural music processing pipelines, with a focus on high-level edits. The dataset has been created with the following applications in mind: training neural models for music editing (e.g., conditional audio generation or manipulation) or music understanding; benchmarking generative audio editing pipelines, particularly in tasks where the goal is to assess how well models preserve or transform musical content based on textual instructions.

\textbf{Societal Impacts}

This dataset can help advance the interpretability and controllability of neural music editing models. By providing aligned audio edits and captions, it could help researchers to: develop models that respond more accurately to user intent in music editing tasks ;
Improve transparency in how models transform musical content, therefore improving human-AI collaboration in creative workflows.
support educational and research efforts in understanding the capabilities and limitations of neural music systems.
Negative Impact/Risks: may reinforce Western-centric biases, could enable low-fidelity or misleading music generation if misused.

\end{document}